 \documentclass[final,authoryear,3p,times,twocolumn]{elsarticle}

\usepackage{subfigure}
\usepackage{graphicx}
 \usepackage{epsfig}
\usepackage{amssymb}

\journal{Journ. of Atm. and Solar-Terrestrial Physics}

\begin{document}

\begin{frontmatter}

\title{Numerical Modeling of Interplanetary Coronal Mass Ejections and Comparison with Heliospheric Images}

\author{N.~Lugaz \& I.~I.~Roussev}
\address{Institute for Astronomy, University of Hawaii--Manoa, Honolulu, HI, USA}
\ead{nlugaz@ifa.hawaii.edu, iroussev@ifa.hawaii.edu}

\begin{abstract}

Interplanetary coronal mass ejections (ICMEs) have complex magnetic and density structures, which is the result of their interaction with the structured solar wind and with previous eruptions. ICMEs are revealed by in situ measurements and in the past five years, through remote-sensing observations by heliospheric imagers. However, to understand and analyze these observations often requires the use of numerical modeling. It is because no instruments can yet provide a simple view of ICMEs in two or three dimensions. Numerical simulations can be used to determine the origin of a complex ejecta observed near Earth, or to analyze the origin, speed and extent of density structures observed remotely. Here, we review and discuss recent efforts to use numerical simulations of ICMEs to investigate the magnetic topology, density structure, energetics and kinematics of ICMEs in the interplanetary space. 

After reviewing existing numerical models of ICMEs, we first focus on numerical modeling in support of the SMEI and {\it STEREO} observations. 3-D simulations can help determining the origins of the fronts observed by SECCHI and SMEI, especially for complex events such as the January 24-25, 2007 CMEs. They can also be used to test different methods to derive ICME properties from remote observations, to predict and explain observational effects, and to understand the deceleration and deformation of ICMEs. In the last part, we focus on the numerical investigation of non-magnetic cloud ejecta. We discuss how simulations are crucial to understand the formation of non-twisted ejecta and the formation of complex ejecta due to the interaction of multiple ICMEs. 

\end{abstract}

\begin{keyword}


\end{keyword}

\end{frontmatter}


\section{Introduction} \label{sec:intro}

Coronal Mass Ejections (CMEs) have been observed remotely from space since the early 1970s with OSO-7 \citep[]{Tousey:1973} and Skylab \citep[]{MacQueen:1974}. Their association with geo-effective ejecta observed in-situ around Earth has been made since the early 1980s \citep[]{Burlaga:1982}. They are most often identified as moving bright fronts in coronagraphic instruments, such as: (1) the K-coronameters (currently Mk-4) of the Mauna Loa Solar Observatory (MLSO) \citep[]{Elmore:2003}; (2) the coronagraphs onboard the {\it Solwind} \citep[]{Sheeley:1980} and the {\it Solar Maximum Mission (SMM)} \citep[]{MacQueen:1980}; (3) the C2 and C3 coronagraphs, part of  the Large Angle Solar COronagraph (LASCO) onboard the {\it Solar Heliospheric Observatory (SoHO)} \citep[]{Brueckner:1995}; and (4) the COR-1 and COR-2 coronagraphs, part of the Sun-Earth Connection Coronal and Heliospheric Investigation \citep[SECCHI, see][]{Howard:2002, Howard:2008} onboard the two {\it Solar-TErrestrial RElations Observatory (STEREO)} spacecraft \citep[]{Kaiser:2008}. These coronagraphs image the corona and the inner heliosphere up to heliospheric distances of about 0.1--0.15~AU (20--30~$R_\odot$). Until the recent launch of {\it Coriolis} containing the Solar Mass Ejection Imager (SMEI) and {\it STEREO}, there had been only a limited number of instruments dedicated to heliospheric observations of CMEs inside of 1~AU, mainly the in-situ and zodiacal polarimeters onboard the two {\it Helios} spacecraft \citep[]{Jackson:1985}. 
\begin{figure*}[t]
\begin{minipage} []{1.0\linewidth}
\begin{center}
\includegraphics[width=6.5cm]{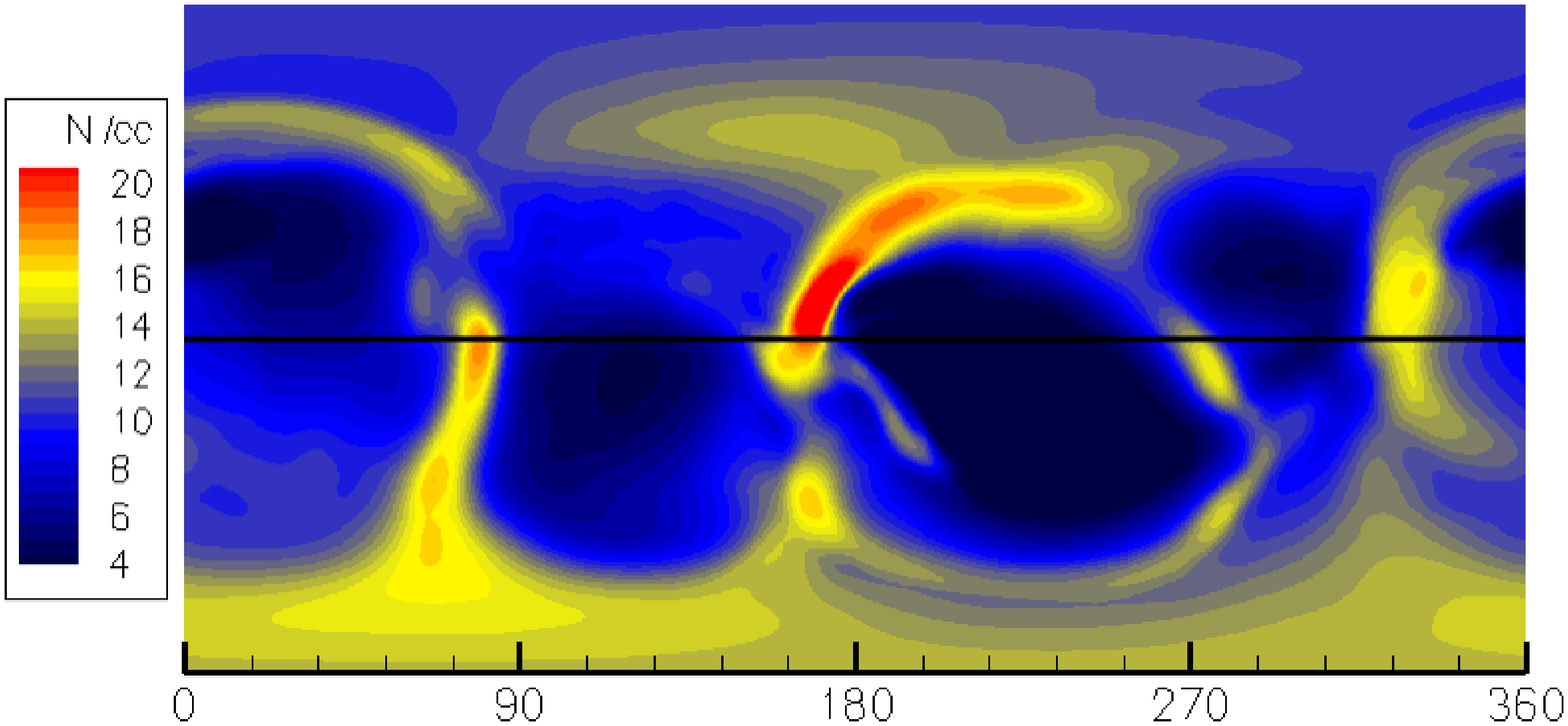}
\includegraphics[width=6.5cm]{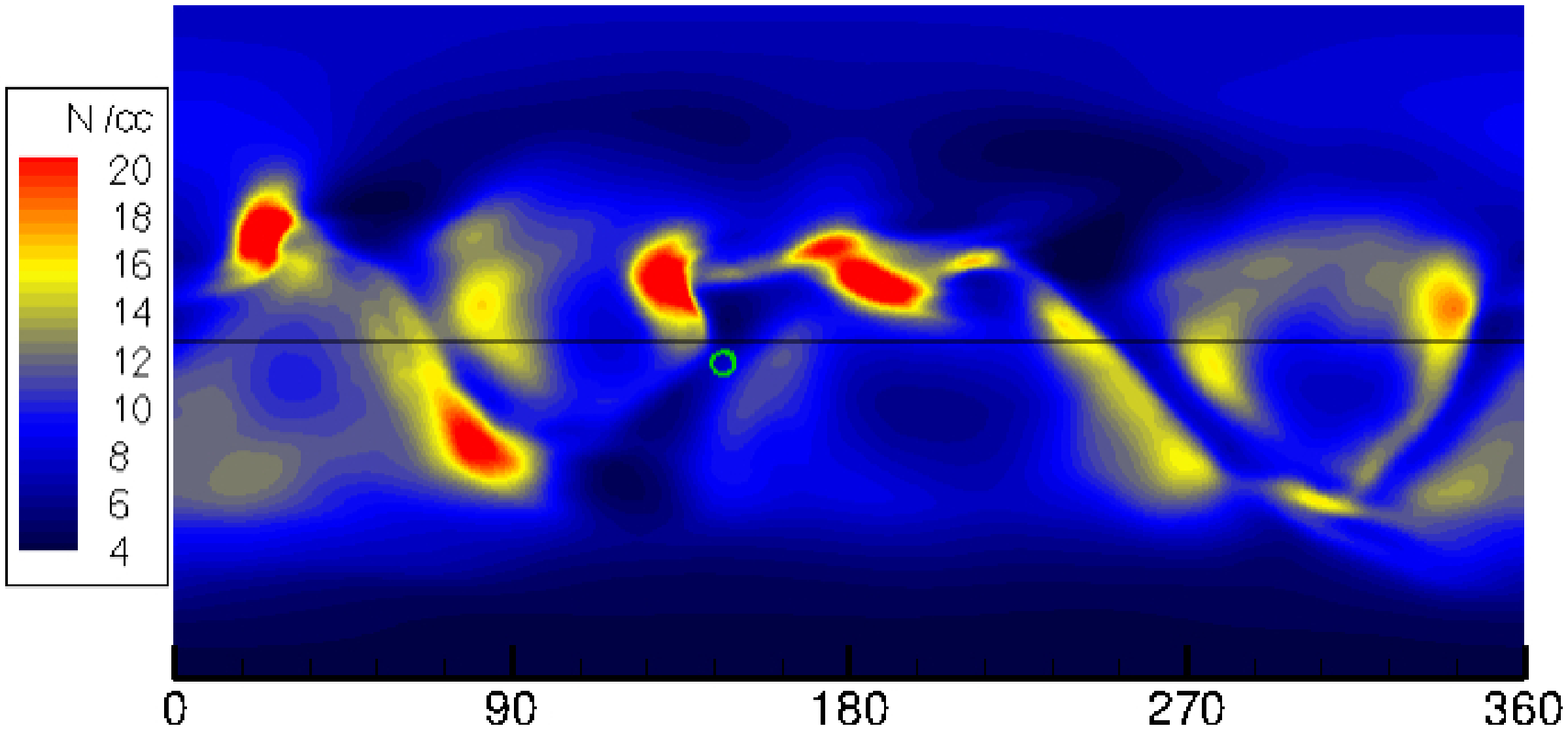}
\includegraphics[width=6.5cm]{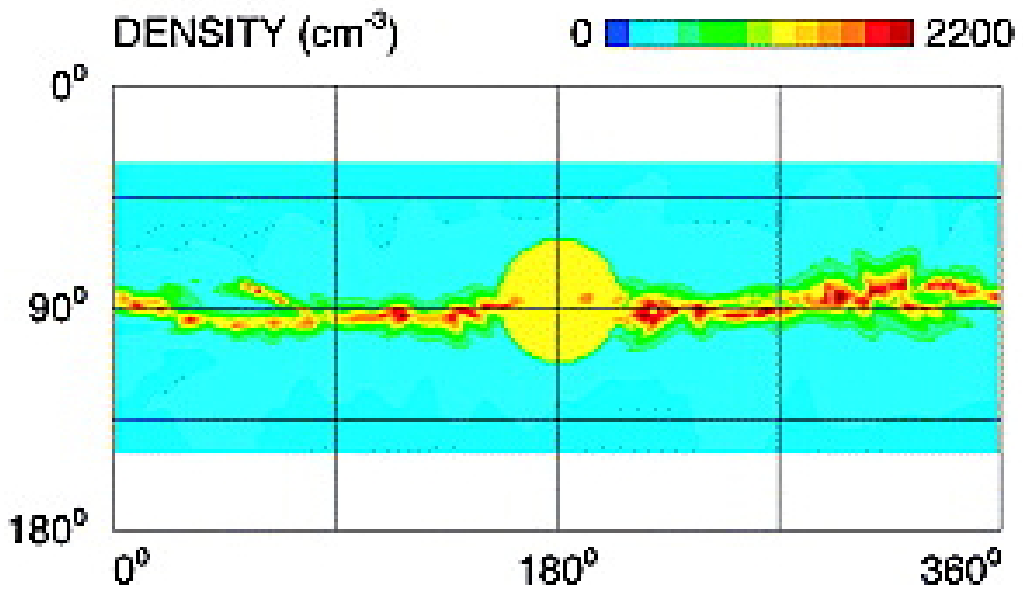}
\end{center}
\end{minipage}\hfill
\caption{Number density from different solar wind models (in cm$^{-3}$). {\it Upper left}: Model of \citet{Roussev:2003b} at 1~AU for the simulation of the August 24, 2002 CME, as reported in \citet{Lugaz:2009a}. {\it Upper right}: Model of \citet{Cohen:2007} at 1~AU for the simulation of the January 24, 2007 CME, as reported in \citet{Lugaz:2009b}. {\it Bottom}: Model of \citet{Odstrcil:2005} at 0.1~AU for the May 12, 1997 CME.} 
\label{fig:sw}
\end{figure*}
To bridge the gap between the first 0.15~AU and Earth's orbit, numerical simulations of interplanetary coronal mass ejections (ICMEs) have been the primary tool to investigate the propagation of ICMEs and their interaction with the interplanetary medium. The complexity of numerical models has been constantly increasing from simple one-dimensional (1-D) hydro-dynamic (HD) models \citep[]{Hundhausen:1969, Gonzalez:2003} or magneto-hydrodynamics (MHD) models \citep[]{Dryer:1976, WuCC:2005} starting in the 1970s to 2-D MHD models in the 1980s and 1990s \citep[]{Dryer:1979, Wu:1983, Vandas:1995, Riley:1997, Holst:2002}. More recently, axi-symmetric 3-D model (so-called 2.5-D) \citep[]{Wu:1999, Cargill:2000, Schmidt:2003} and fully 3-D MHD models \citep[]{Han:1988,Gombosi:2000, Groth:2000, Riley:2001, Manchester:2004b, Odstrcil:2004, Hayashi:2005, Jacobs:2006, Nakamizo:2009, Kataoka:2009} have been developed and used to study ICMEs. Now, we can for the first time constrain and test these models with heliospheric remote observations of ICMEs. But, the use of simulations goes beyond reproducing observations. Numerical investigations of the propagation of ICMEs have been central to further our understanding of  the deceleration of ICMEs in the heliosphere, the interaction of ICMEs with the bimodal solar wind and their associated deformation, the association of magnetic clouds observed in situ with CMEs and the interaction of multiple ICMEs to form complex ejecta. Most of these physical processes can not be investigated with the existing observations and numerical simulations are required. In this review, we summarize past and present efforts to simulate ICMEs in section 2, and we present some recent progresses regarding the simulation of ICMEs in support of observations with the SMEI instrument and the {\it STEREO} mission in section 3. In section 4, we report on using simulations to test observational methods and we explore some of the major advancements in our understanding of complex ejecta observed near Earth in section 5. The final conclusions are drawn in section 6.

\section{Numerical Simulations of ICMEs}

Heliospheric numerical models follow two main approaches: (1) using an inner boundary past the critical point (where the fast Mach number is unity, usually beyond 0.1~AU), or (2) coupling the heliospheric model self-consistently with a solar coronal model. Both approaches have advantages and drawbacks. Mainly, because the fast magnetosonic speed in the heliosphere is much less than that in the lower corona (usually by two orders of magnitude), the timestep in a heliospheric model is much greater than that in a solar coronal model (a fraction of a minute to a minute in an heliospheric model compared to about a tenth of a second in a coronal model). Therefore coupling an heliospheric and a solar coronal models (approach 2) is much more time intensive than just running an heliospheric model (approach 1) and it can only be implemented on relatively large machines (64--256 CPUs). Because of the difference in timesteps, a heliospheric model runs much faster than real time even on a small number of CPUs (4--32 CPUs). A typical run requires only a few hundred of CPU hours for a 0.1 AU to 1 AU propagation. A coupled simulation can only run close to realtime on much larger machines (128--256 CPUs) and a typical run is 5,000-50,000 CPU hours for a Sun to Earth propagation, depending on the resolution. However, determining the inner boundary conditions in approach 1 is a complex problem. It is especially true for the initiation of 3-D CMEs around 0.1~AU with a realistic density structure, magnetic structure and velocity field. By comparison, initiating a CME on the solar surface (approach 2) yields complex and realistic structures at 0.1~AU. Moreover, approach 2 is the only one which can be used to study the interplanetary consequences of different initiation mechanisms \citep[see review by][in this volume]{Poedts:2010}. Here, we review some of the solar wind and CME models used for heliospheric numerical studies.

\begin{figure*}[t]
\begin{minipage} []{1.0\linewidth}
\begin{center}
\includegraphics[trim = 270 20 0 0 , clip = true, width=6.5cm]{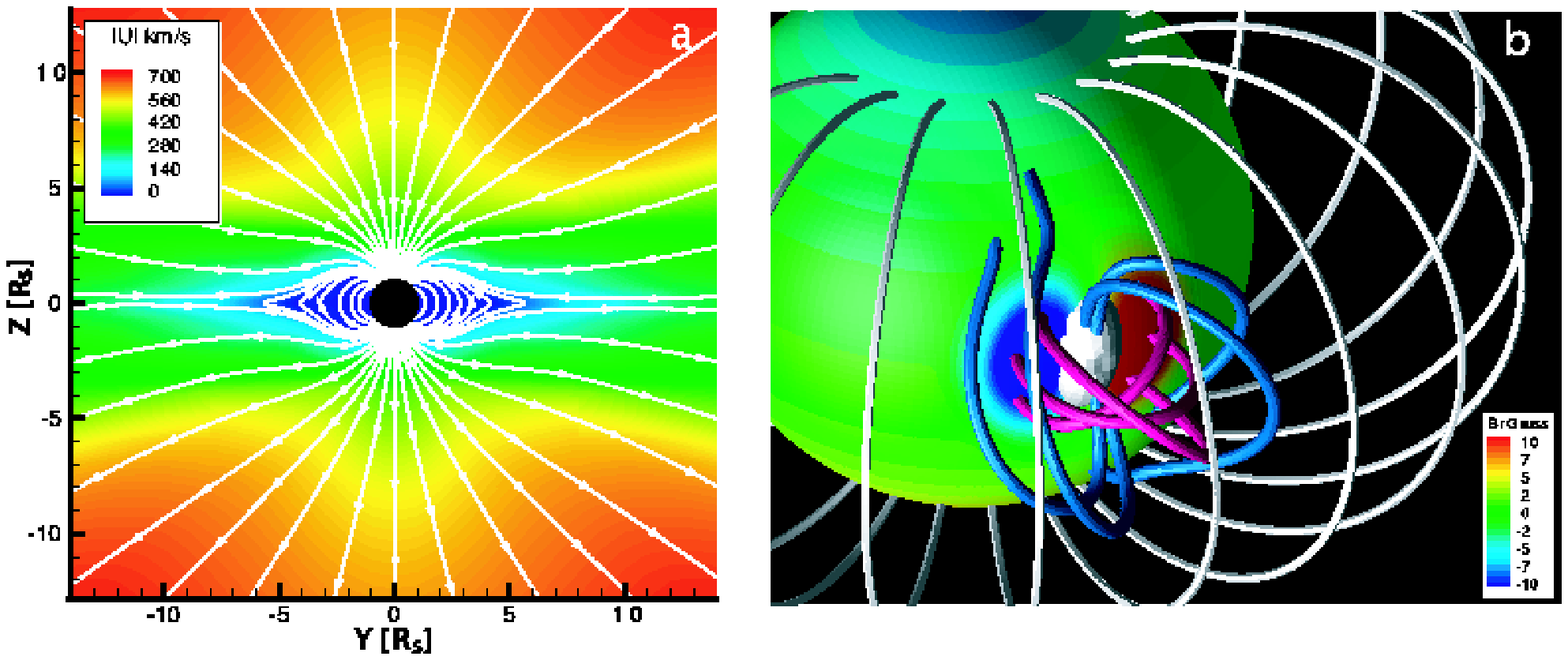}
\includegraphics[width=6.5cm]{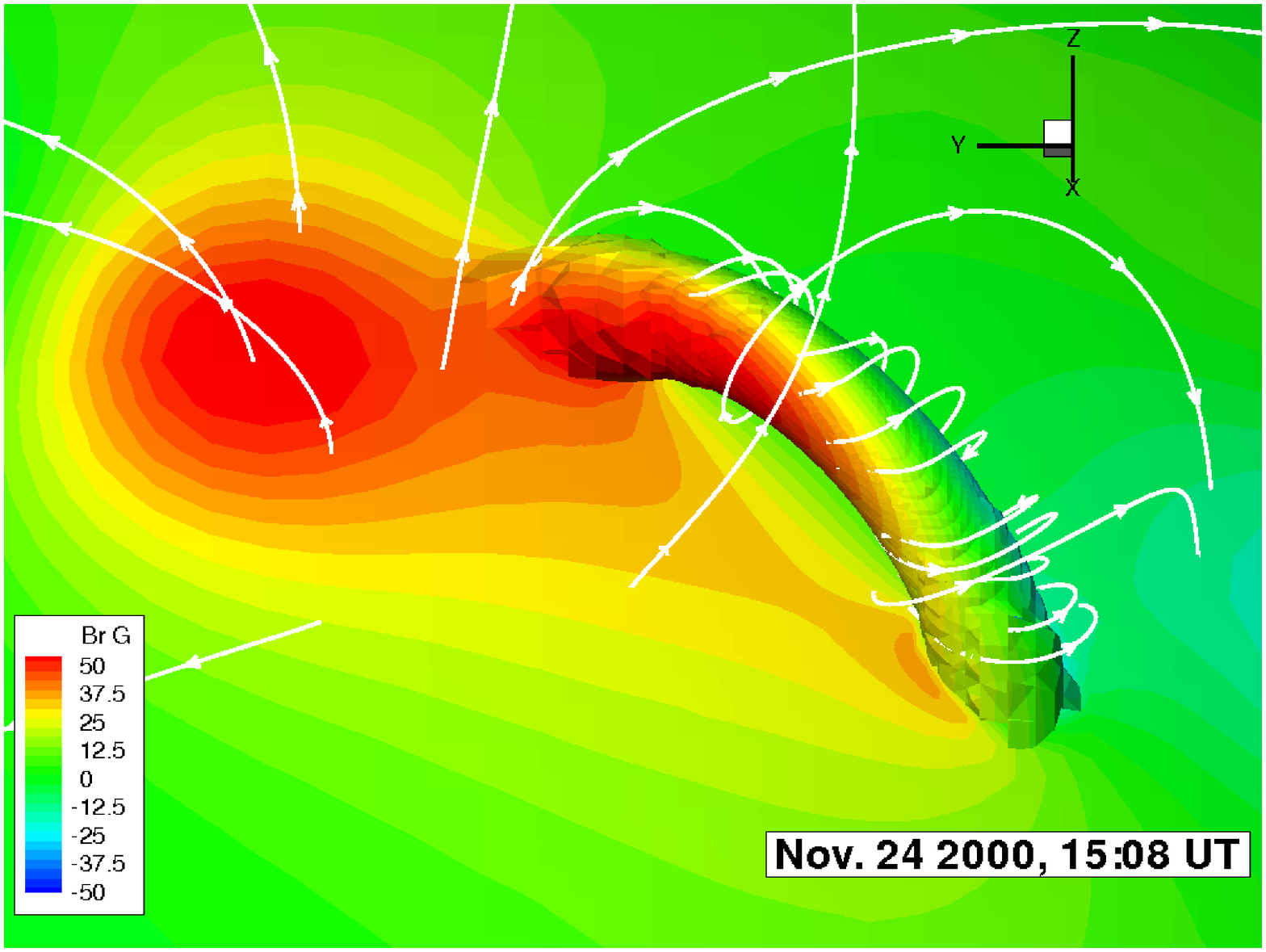}
\includegraphics[width=6.5cm]{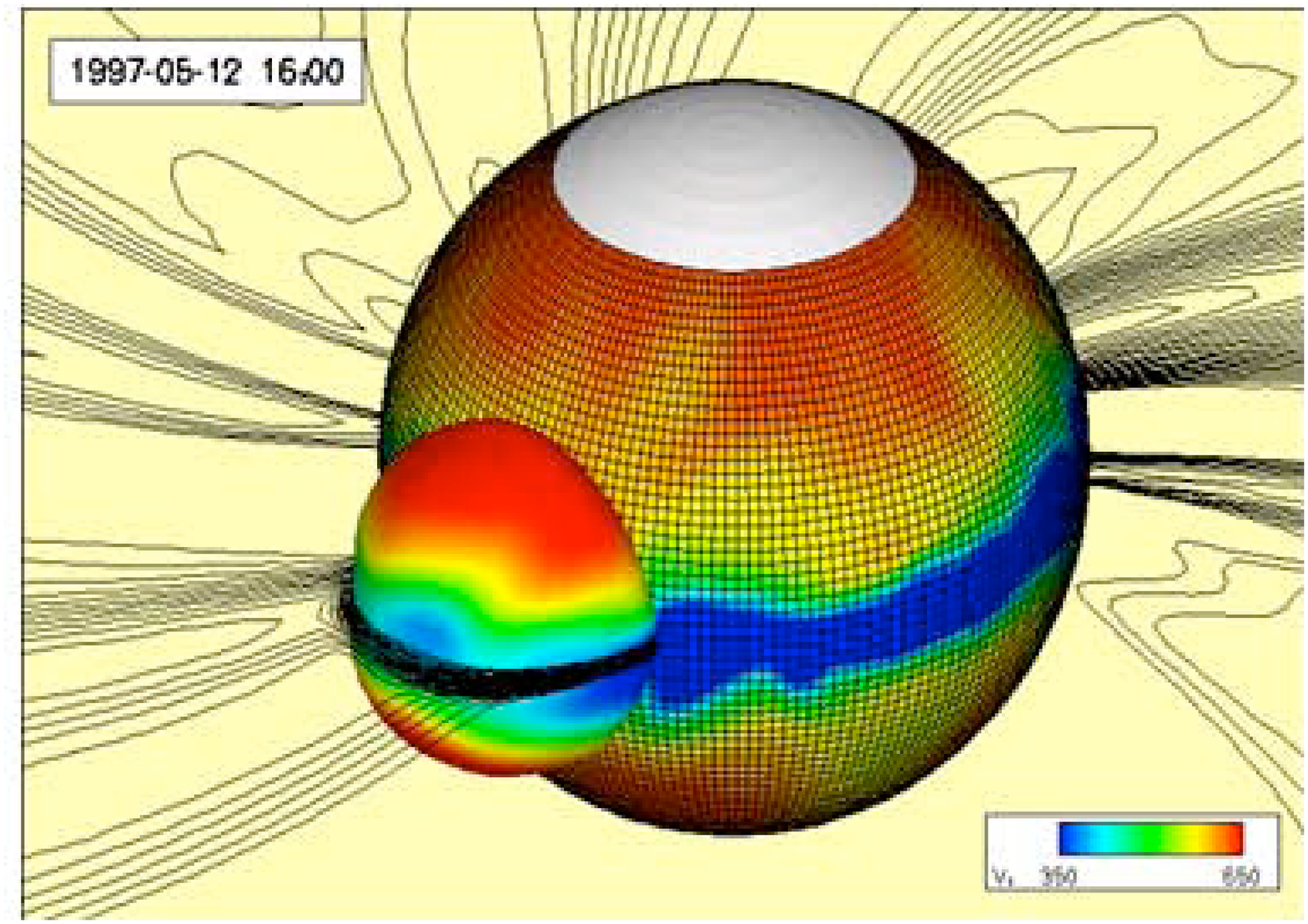}
\end{center}
\end{minipage}\hfill
\caption{{\it Upper left}: Spheromak-type model for coronal mass ejections \citep[]{Gibson:2000} as used in the model of \citet{Manchester:2004b}. {\it Upper right}: Semi-circular flux rope model of \citet{Roussev:2003a} used in the simulation of \citet{Lugaz:2007} and based on the Titov-Demoulin flux rope \citep[]{Titov:1999}. The image is zoomed on active region 9236 and shows the solar surface and an isosurface of density color-coded with the radial magnetic field. {3-D} magnetic field lines are drawn in white. {\it Bottom}: Coronal Mass Ejection \citep[cone model, see][]{Zhao:2002, Odstrcil:2004} used in the model of \citet{Odstrcil:2009}} 
\label{fig:NotTD}
\end{figure*}

\subsection{Solar Wind Models}

The simplest numerical models of solar wind use analytical expressions for the solar wind speed, which increases approximatively like the logarithmic of the distance, and for the solar wind density which decreases  $\propto r^{-2}$ in the heliosphere. The next level of complexity is to define an axisymmetric bimodal solar wind, reflecting solar minimum conditions, where the slow solar wind has a speed of 300--400 km~s$^{-1}$ and a density of 5--15~cm$^{-3}$ at 1~AU and the fast wind a speed of 700--900 km~s$^{-1}$ and a density of 1--5~cm$^{-3}$. Such a solar wind can be set up using both approaches described above. It has been done in \citet{Han:1988} or \citet{Odstrcil:1999} with an inner boundary at 0.1~AU, and in \citet{Wu:1999}, \citet{Groth:2000} and \citet{Gombosi:2000} with an inner boundary at the solar surface. This class of models is particularly useful to study ICME deformation associated with the bimodal solar wind (see section 3.1). Finally, the most advanced class of models uses solar observations, typically magnetogram, as input to create a full 3-D solar wind with realistic speeds and densities. Most models now use the Wang-Sheeley-Arge relation \citep[WSA model, see][]{Wang:1990, Arge:2000} between flux tube expansion and speed at 1~AU. It can be used with a boundary at 0.1~AU \citep[]{Odstrcil:2005} or at the solar surface \citep[]{Cohen:2007, Oran:2010}.  Other models such as those proposed by \citet{Usmanov:1993b}, \citet{Roussev:2003a} and  \citet{Odstrcil:2004} involve empirical terms inspired by physical processes (wave-particle interaction) to create a distribution similar to that obtained from the WSA model. Figure~\ref{fig:sw} shows three examples of steady-state solar wind density distributions used in recent simulations. The top panels, left and right,  show results from the solar wind models of \citet{Roussev:2003a} and \citet{Cohen:2007} based on a coupled corona-heliosphere code.  The bottom panel shows the result from the heliospheric model of \citet{Odstrcil:2005}. All three models have a higher-density current sheet and lower density regions corresponding to the poles and coronal holes. The values obtained at 1 AU for the density between 3 and 20 cm$^{-3}$ are typical to what is observed in-situ.

\subsection{Coronal Mass Ejection Models}
Most solar wind models are very similar in both approaches, i.e. independently of the height where the inner boundary is set. In general, the same is not true for CME models. 
One exception is CME ``models'' which consist in adding a pressure or velocity pulse at the inner boundary. This simplest type of models has been used for simulations starting close to the solar surface \citep[]{Usmanov:1995, Groth:2000, Chane:2006} and past the fast magnetosonic point \citep[]{Odstrcil:2002b}.
CME models initiated at the solar surface have been recently reviewed in \citet{Roussev:2006} and in \citet{Poedts:2010}. To summarize, there are two main classes of models for CMEs initiated at the solar surface: with or without a pre-existing flux rope. A pre-existing flux rope can be superposed out of equilibrium to the steady-state surface \citep[]{Lugaz:2007, Toth:2007}, can emerge from below the photosphere \citep[]{Amari:2004, Manchester:2004c, Archontis:2008} or can be superposed in equilibrium and erupt due to flux cancellation or boundary motions \citep[]{Wu:1999, Chen:2000, Roussev:2003b}. In the absence of a pre-exsiting flux rope, boundary motions (shearing and/or converging motions) can result in a loss of equilibrium and the formation of a flux rope \citep[]{Mikic:1995, Antiochos:1999, Amari:2000, Linker:2003, Lynch:2005, Holst:2007, Aulanier:2010}. 

For heliospheric models with an inner boundary beyond the critical point, these physical initiation mechanisms are very hard (for models using pre-existing out-of-equilibrium flux ropes) or even impossible (for models invoking boundary motions) to adapt. An alternative way, more realistic than simply using a pressure pulse, is to use geometrical models of CMEs. It is often done with a cone model \citep[]{Zhao:2002, Xie:2004, Xie:2006b}, which can be fitted to the particular characteristics of the studied event \citep[]{Odstrcil:2004}. Even using this model, however, it is hard to define the three-dimensional distribution of the density and the magnetic field as well as the velocity field. Often, the ICME is treated as a purely hydrodynamical perturbation (no magnetic field) and it is set with a uniform density and velocity inside the cone. Recently, it was improved to better reflect the flux rope shape of CMEs,  to use information obtained from stereoscopic fitting \citep[]{Odstrcil:2009} and also to include non-uniform variation of the density and velocity \citep[]{Hayashi:2006}. Three examples of CME models used in recent heliospheric simulations are shown in Figure~\ref{fig:NotTD}, illustrating two types of flux-rope models at the solar surface and one example of a cone model at 0.1~AU. The top panels show two flux ropes: a spheromak-type flux rope developed by \citet{Gibson:2000} (left) and a semi-circular flux rope developed by \citet{Titov:1999} (right), both added onto the solar surface. In both figures, the color shows the radial magnetic field strength in Gauss, illustrating how the flux ropes are added onto active regions. The bottom panel shows the cone model of \citet{Zhao:2002} at the inner boundary of the code. The color shows the radial velocity, which illustrates how this ejection is added across the slow solar wind and extends into the fast solar wind from its onset.

\begin{figure*}[ht*]
\begin{minipage} []{1.0\linewidth}
\begin{center}
\subfigure{\includegraphics*[width=6.5cm]{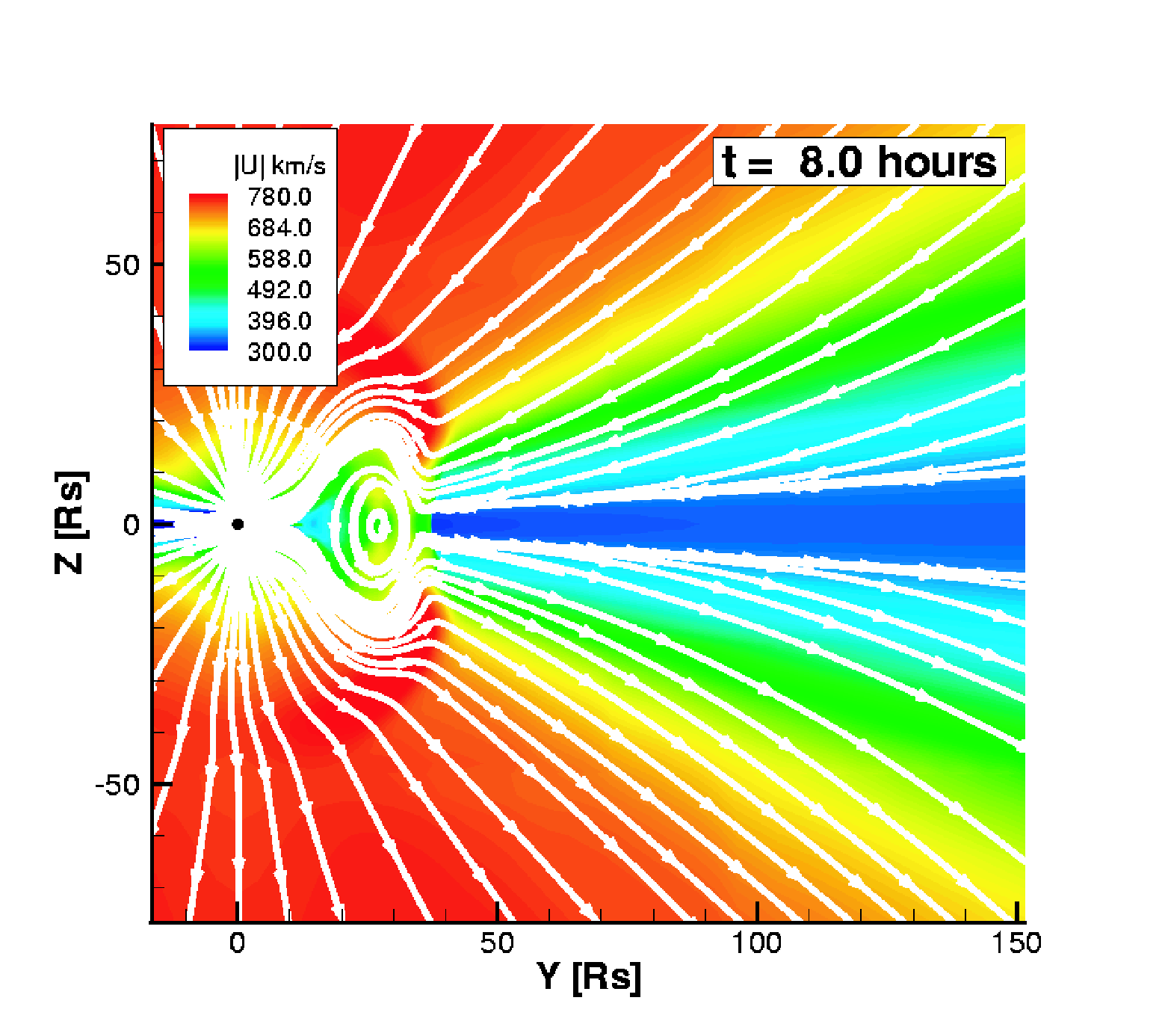}}
\subfigure{\includegraphics*[width=6.5cm]{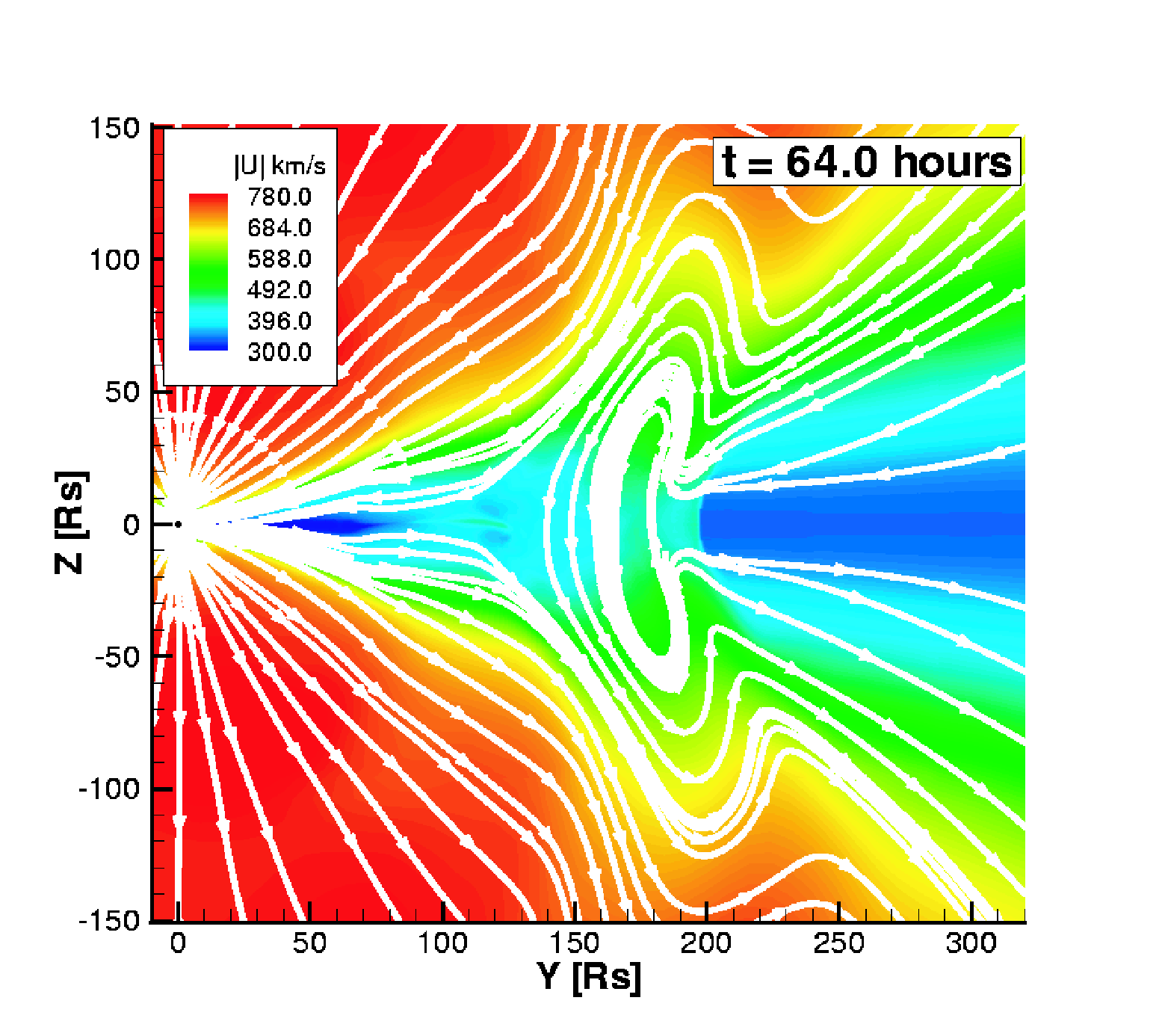}}\\
\subfigure{\includegraphics*[width=4.8cm]{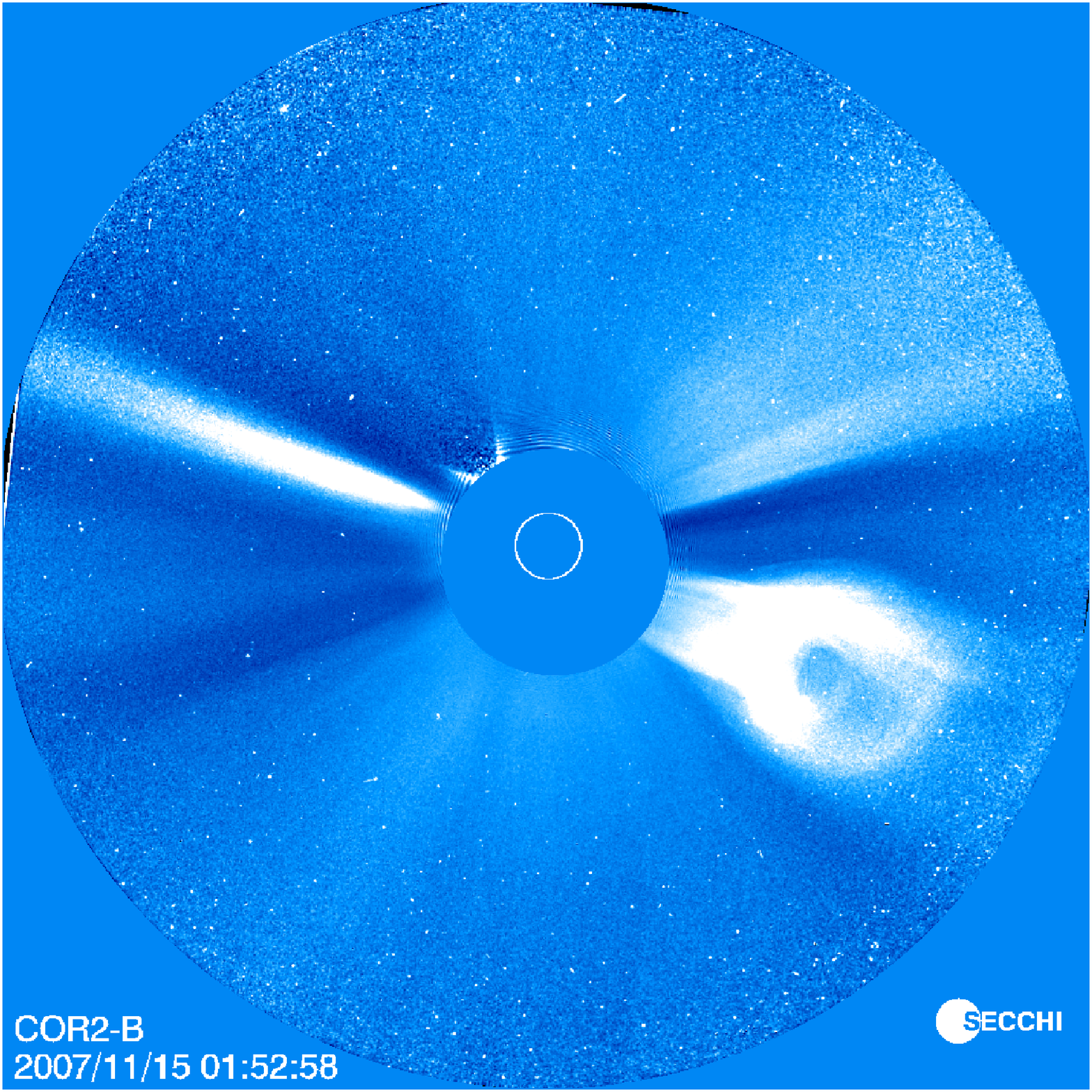}}\hspace{2cm}
\subfigure{\includegraphics*[width=4.8cm]{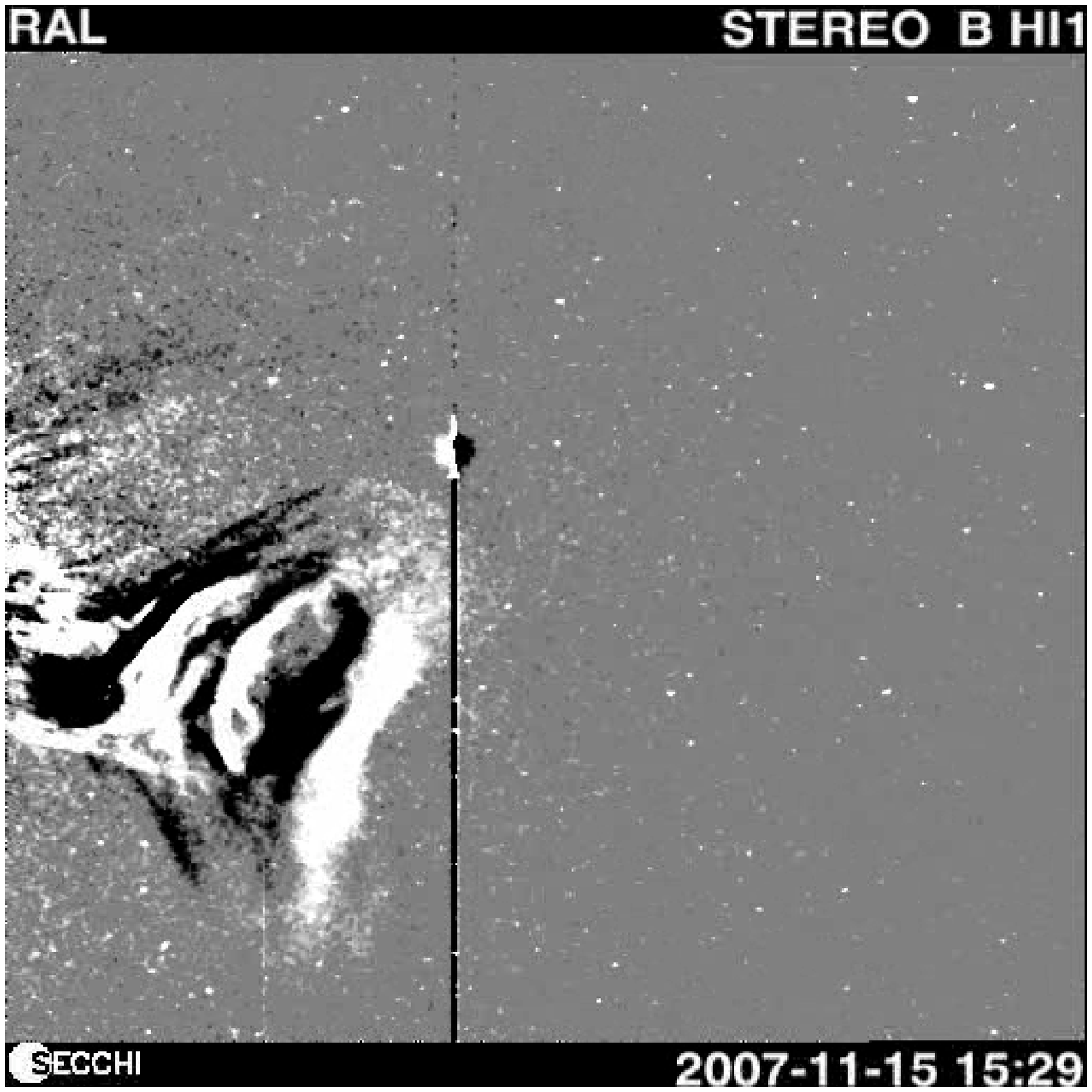}}
\end{center}
\end{minipage}\hfill
\caption{{\it Top}: CME interaction with the solar wind. Solid white lines streamlines display magnetic ``streamlines" superimposed upon a color image of the velocity magnitude on the $y-z$ plane ({\it from \citet{Manchester:2004b}}). {\it Bottom}: View of the November 14, 2007 CME reported by \citet{Savani:2010} at two different times in {\it STEREO-B}/COR-2 ({\it left}) and HI-1 ({\it right }) fields-of-view.}
\label{fig:distortion}
\end{figure*}

\subsection{Reproducing Observations With Numerical Models}

It is straightforward to extract a synthetic satellite file from a time-dependent numerical model, knowing the spacecraft orbit. Such in situ measurements of the plasma parameters have been compared to generic CMEs measured in situ \citep[]{Riley:2003, Manchester:2004b} and to specific events, recently the Halloween storm of November 2003 \citep[]{Toth:2005, WuCC:2005}, the November 24-25, 2000 CMEs \citep[]{Lugaz:2007}, the May 1997 CME \citep[]{Odstrcil:2004, Odstrcil:2005, WuCC:2007, Cohen:2008a}. Additionally, white-light images synthesis has been done in heliospheric codes since the mid 2000s \citep[]{Lugaz:2005a, Odstrcil:2005, Sun:2008}, prior the launch of the {\it STEREO} mission. Some predictions of numerical models in support of the {\it STEREO} mission have been summarized in \citet{Aschwanden:2008}. The procedure to create these images is less straightforward than that to extract satellite data, but still relatively easy. Coronagraphs and heliospheric imagers measure the Thomson-scattered light, which only depends on geometrical factors and the electron density along a given line-of-sight \citep[]{Billings:1966, Hundhausen:1993}. In a 3-D numerical simulation, the plasma density is known everywhere and a line integration can be used to calculate the Thomson integral. Good agreement was found between observed and synthetic coronagraphic line-of-sight images \citep[e.g., see][]{Linker:1999, Lugaz:2007} as well as with eclipse images \citep[e.g., see][]{Mikic:1999, Rusin:2010}. Some recent works \citep[]{Manchester:2008,Lugaz:2009b} have focused on making the synthetic procedure of creating coronagraphic images as close as possible from the real procedure, either by accounting for the F-corona \citep[the light scattered by dust particles in][]{Manchester:2008} or by creating a background image for the MHD simulation using the procedure developed for the real images \citep[]{Lugaz:2009b}.

\section{Support for Heliospheric Remote Imaging}

\subsection{CME Deformation and Deceleration}

As the only way to study the 3-D properties of ICMEs, numerical simulations have been used to make predictions regarding the propagation of ICMEs in the interplanetary space. Numerical investigations have also helped determining which type of forces acts on ICMEs \citep[e.g., see][]{Chen:1996, Tappin:2006}. ICMEs appear to be decelerated in the heliosphere by a drag-type force (see below) but there is also some evidence of a driving force, as some ICMEs accelerate or do not decelerate much in the heliosphere. \citet{Chen:1996} proposed that it is caused by the Lorentz force acting on the ICME. Alternatively, the presence of a fast stream catching up with an ICME is expected to accelerate it \citep[]{Gopalswamy:2009} and it could be the reason why some CMEs do not decelerate. Often, observations are not accurate enough to determine whether the ICMEs decelerate or not, and numerical simulations must then be used.

In \citet{Lugaz:2005a}, using the simulation of \citet{Manchester:2004b}, we studied the evolution of the CME mass as the CME propagates in the interplanetary space. We found that the mass increases by a factor of 5 from 1~$R_\odot$ to 1~AU. This mass increase is associated with the solar wind material piled-up and compressed by the shock wave and incorporated into the CME. It is important to know this factor because, until the launch of {\it STEREO} and {\it Coriolis}, CME masses could only be determined close to the Sun \citep[]{Vourlidas:2002} but mass appears to have some effects on the associated geo-magnetic storms \citep[]{Farrugia:2006}. Numerical simulations are the easiest way to study how the CME mass changes as a CME propagates. 

A number of studies focused on the cause of the observed ICME heliospheric deceleration \citep[e.g., see][]{Riley:1997, Cargill:2004, Vrsnak:2004}. Two physical models proposed to explain this deceleration are: (1) the ``snow-plow'' model where the momentum of the system is conserved and mass increase is associated with deceleration \citep[]{Tappin:2006}, and (2) the heliospheric drag acting on the CME to match its speed to that of the solar wind (without being associated with mass increase).  The drag force can be written $C_D (V_{CME} - V_{sw}) |V_{CME} - V_{sw}|$, where $V_{CME}$ and $V_{sw}$ are the CME and solar wind speeds, respectively. Estimations of the drag coefficient, $C_D$, have been made based on numerical simulations in \citet{Vrsnak:2001}, \citet{Cargill:2004} and \citet{Vrsnak:2010}, among other works, reporting values for the drag coefficient between 1 and 10~$AU^{-1}$ (except for extremely slow ICMEs), depending on the initial speed and mass of the ICMEs, but remaining relatively constant for a given CME from the solar corona to 1~AU.

In all numerical studies of CMEs which include an idealized bimodal solar wind, it was found that wide ICMEs are significantly deformed as they propagate \citep[]{Riley:1997, Cargill:2002}. It is due to the large drag force encountered near the current sheet, where the density is high and the solar wind speed is low. In contrast, ICMEs encounter a weaker drag force at high latitudes where the density is lower and the solar wind speed faster. The difference in the magnitude of the drag force can result in a concave-outward deformation of the ICME and its shock wave. 
Before the launch of the {\it STEREO} mission, testing findings from heliospheric numerical studies required statistical analyses using in-situ measurements. Such an investigation by \citet{Liu:2008}  found some signs that magnetic clouds during solar minimum conditions have concave-outward curvature, but it also found that the deflection flow ahead of the cloud was equatorward, which is the opposite behavior to that predicted by numerical models \citep[for example, see][]{Manchester:2005}. No sign of concave-outward indentation was found in another study by \citet{Owens:2004}. While ICMEs appear deformed and flattened in HI-1 field-of-view (see Figure~\ref{fig:distortion}), there has been only one clear instance of a change of indentation of the ICME leading edge in SECCHI field-of-view \citep[]{Savani:2010}. V-shape density enhancements are often observed at the back of CMEs \citep[]{Kahler:2007} both by SMEI and SECCHI but this appears to be a different phenomenon. The deformation of a magnetic cloud and a ICME leading edge requires a background solar wind which is at the same time steady and close to axi-symmetric. The presence of Corotating Interaction Regions (CIRs), dense streams and previous eruptions do not create conditions optimal to strongly deform ICMEs. The current solar minimum has had significantly more CIRs than the previous minimum \citep[]{Mason:2009}, which might explain why such indentation changes have not been observed more often by SECCHI. It is also worth noting that recent simulations using non-axisymmetric solar wind models and/or multiple CMEs predict a flattening of the CME front but not necessarily a change of its indentation \citep[see simulations in][]{Toth:2007, Lugaz:2007, Lugaz:2009b, Webb:2009}.

\subsection{Case Study of the January 24-25 CMEs}

\begin{figure}[ht*] 
\centerline{\includegraphics*[trim = 10 318 20 0 , clip = true, width=8cm]{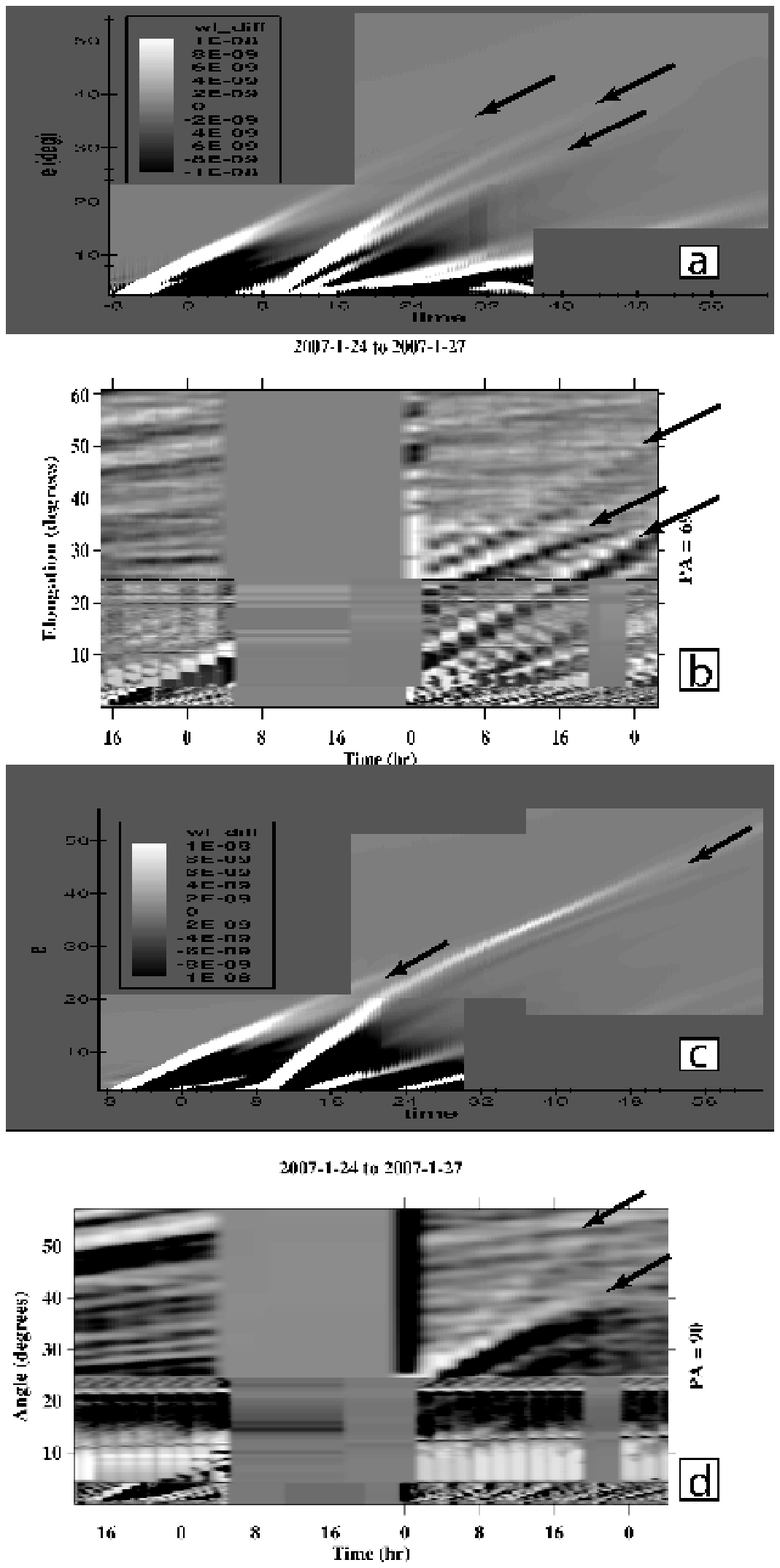}}
\caption{Comparison of synthetic (a) and real (b) J-maps for PA~69$^\circ$, the center of the HIs image. The first tick on the $x$ axis corresponds to 16:00UT on January 24. Arrows point to similar tracks observed in the synthetic and real J-maps.}
\label{fig:Jmap}
\end{figure}

After the launch of the two {\it STEREO} spacecraft in 2006 October 26, one of the first CME events observed by the SECCHI suite was the two eruptions of January 24 and 25, 2007 \citep[]{Harrison:2008, Harrison:2009}. On 2007 January 24 at 14:03UT, {\it STEREO-A}/COR1 observed an eruption whose source region was behind the eastern limb of the Sun, with a speed of about 600-750~km~s$^{-1}$. At 06:43UT on January 25, COR1 observed a second eruption also originating from behind the Sun and with a speed of about 1200-1400~km~s$^{-1}$. 
There was a {\it STEREO-A}/SECCHI data gap for 20 hours starting at 04:02UT on January 25, and a similar data gap for {\it STEREO-B}/SECCHI starting at 09:02UT on January 25. Multiple fronts were detected in HI-2 after the data gap on January 26: the brightest one propagated from an elongation angle of 24.7$^\circ$ at 02:01UT to an angle of 32.5$^\circ$ at 18:01UT (see Figure~\ref{fig:Jmap}b). There are a number of reasons to numerically study this particular series of events: (1) the 20-hour gap, which complicates the analysis of the data; (2) the fact that this is the first and still one of the best examples of CME-CME interaction observed by SECCHI; and, (3) the fact that the January 25, 2007 was the fastest CME observed by SECCHI between the launch of {\it STEREO} and 2010. 

3-D simulations of these eruptions were reported in \citet{Lugaz:2008b}, \citet{Lugaz:2009b}, \citet{Webb:2009} and \citet{Odstrcil:2009}. Here, we present a comparison between simulations and observations as well as important results from these simulations. Performing numerical investigations of specific events has several goals. First, it is the only true and appropriate test and validation for numerical models. Beyond simply reproducing data, simulating real events is also one of the best ways to study the influence of the solar wind on ICMEs, by including not only an idealized bimodal solar wind, but a realistic 3-D solar wind, with dense streams, fast streams and CIRs. In the case of successive CMEs, it is also important to move away from idealized simulations such as those of \citet{Lugaz:2005b}, \citet{Xiong:2006} and \citet{Xiong:2007} to integrate the complexity of real events and study the interaction between ICMEs of different speeds and directions and the structured solar wind.

The simulation of \citet{Lugaz:2009b} used the Space Weather Modeling Framework \citep[SWMF, see][]{Toth:2005} and coupled a solar coronal model \citep[]{Cohen:2007} to an heliospheric model. The CMEs were initiated with out-of-equilibrium flux ropes \citep[]{Titov:1999, Roussev:2003b} added onto the steady-state corona in the same active region behind the Sun's eastern limb, in a way similar to that shown in the top right panel of Figure~\ref{fig:NotTD}. The two CMEs were initiated 16.67 hours apart and the second CME reaches the back of the first CME at approximatively 12:00UT on January 25. Later on, the overtaking shock wave exits the magnetic ejecta, and it then reaches the back of the dense sheath associated with the first CME around 18:00UT on January 25. The two shocks merge at around 06:00UT on January 26 (see Figure~5 for 2-D cuts of the simulation at these different times). 

Figure~\ref{fig:Jmap} shows a comparison between synthetic and real time-elongation maps (J-maps) made from radial strips along the same position angle (PA): PA~69$^\circ$, the apparent central position angle of the SECCHI instruments at the time of these ejections.
The real J-map (Figure~4b) at this PA is complicated and hard to analyze due to the lack of observations on January 25. There are three tracks visible in HI-2 from the beginning of  January 26 to about 14:00UT when a fourth track enters HI-2's field-of-view. From analytical models taking into consideration the initial speeds of the two ejections, it was expected that they would have merged by the beginning of January 26 \citep[]{Webb:2009}. The presence of three tracks in the HI-2 field-of-view is therefore unexpected and hard to explain from the observations only.
To fill the data gap and to understand the origin of the three tracks, we analyze the real J-map with the help of the J-map created from the simulation (Figure~4a).
In the synthetic J-map at PA~69$^\circ$, there are also three main distinct tracks (marked by black arrows). They correspond to: the first ejection, the second ejection and a dense stream from left to right, respectively. The dense stream is first compressed by the first CME and can be seen around 08:00UT on January 25 behind the first CME in Figure~\ref{fig:Jmap}a. At 04:00UT on January 26, there are three fronts past 20$^\circ$, the brightest corresponding to the second CME. 

 \begin{figure*}[ht] 
 \begin{center}
\subfigure{\includegraphics*[width=9.2cm]{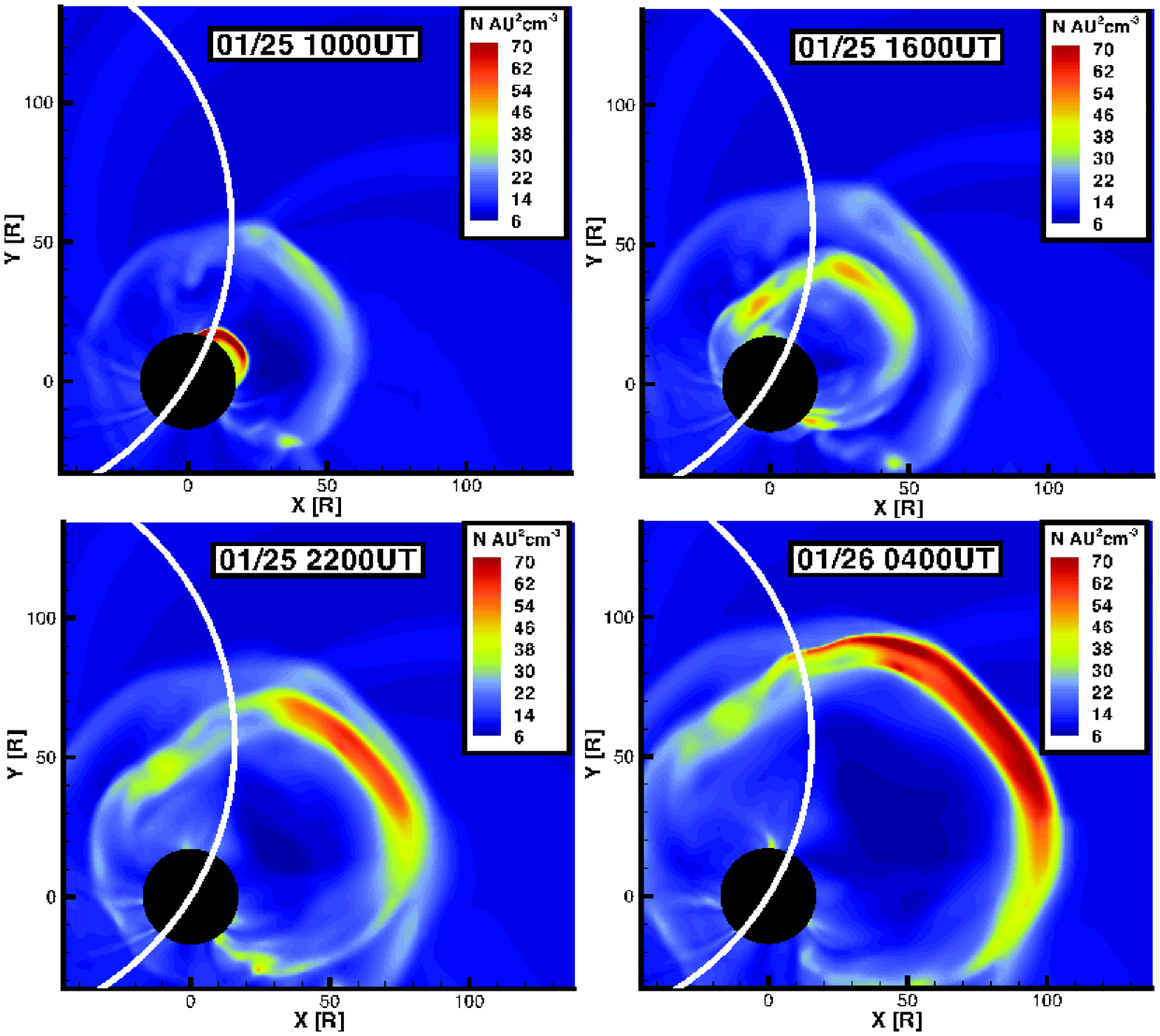}}
\subfigure{\includegraphics*[width=4.2cm]{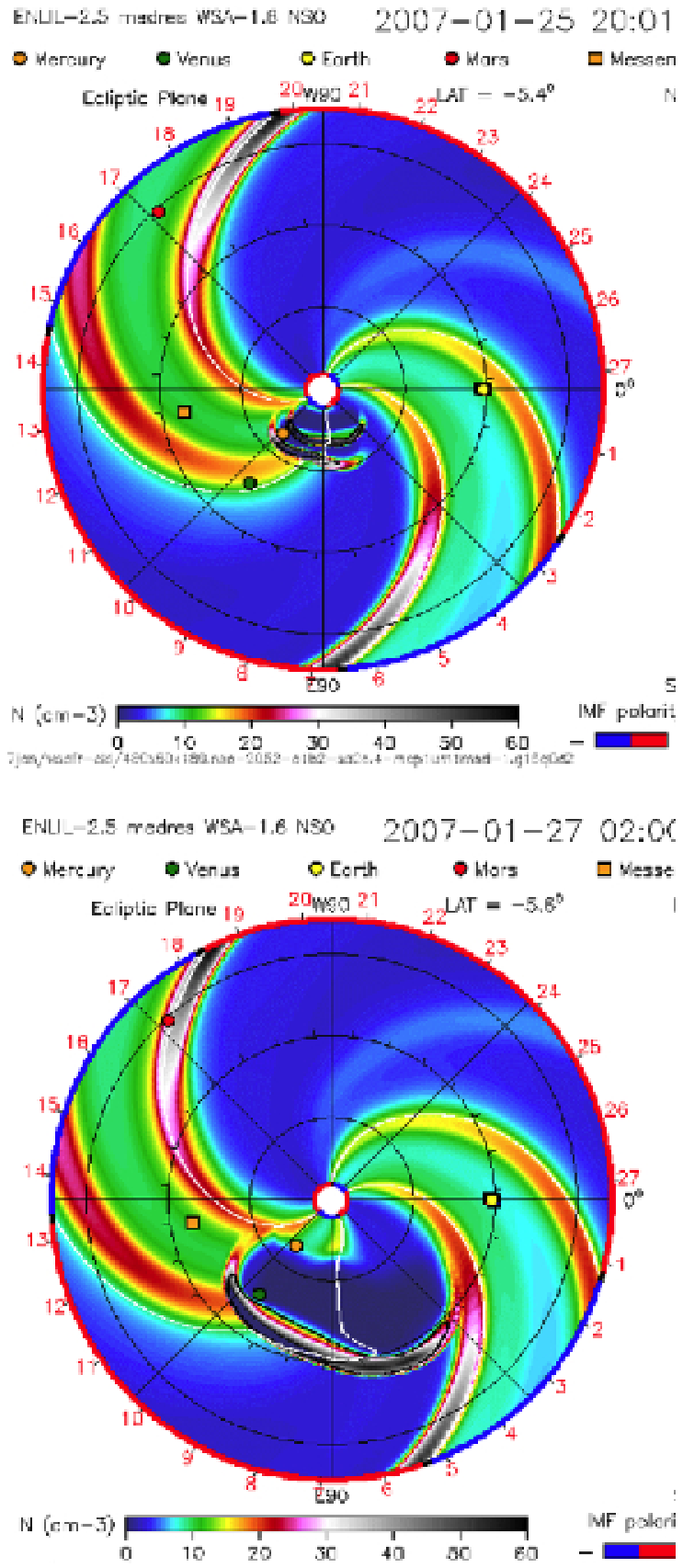}}
 \caption{{\it Left and Middle}: View of the January 24 and 25, 2007 CMEs in the ecliptic plane ($x,y$ plane) from the solar north ($+z$ direction) at four different times prior and during their interaction showing the density scaled by 1/R$^2$ as simulated with the SWMF in \citet{Lugaz:2009b}. The white disk's radius is 16 $R_\odot$ and the black circle is the approximate projection of the Thomson sphere (relative to {\it STEREO}-A) onto the plane of the image. Earth's position is approximatively $(-179, 110, -20)R_{\odot}$ in this coordinate system. {\it Right}: View of the same CMEs from the solar north at two different times before and after the interaction as simulated with ENLIL in \citet{Webb:2009}. Earth's position is marked with the yellow circle at longitude 0$^\circ$. }
 \end{center}
 \label{fig:four}
\end{figure*}

We believe that the three tracks in the real image correspond to the fronts identified in the synthetic image. According to our analysis, the third track corresponds to the dense stream as explained in \citet{Lugaz:2008b}. It is visible before the interaction in the top left panel of Figure~5 at ($X=50~R_\odot$, $Y=70~R_\odot$) and after the interaction in the bottom middle panel of Figure~5 at ($X=20~R_\odot$, $Y=60~R_\odot$). According to our 3-D simulation, this stream is present in the heliosphere prior to the CME passage, with a density about 50$\%$ larger than the average density at the same heliospheric distance.  The simulation shows how the stream is compressed by the two CMEs, and due to the co-rotation, it enters the field-of-view of HI-2 in January 26.  The same dense stream is also clearly visible at both times in the simulation of \citet{Odstrcil:2009}.

The simulation of \citet{Odstrcil:2009} uses ENLIL and starts at 0.1~AU with boundary conditions given by the WSA model. The results are shown in the right column of Figure~5. For this simulation, the authors use a cone model input for the CMEs, which is purely hydrodynamic (no enhanced magnetic field inside the ejecta). This simulation also predicts the merging of the two ejections and the existence of a third dense front due to a corotating dense stream, but it predicts that this CIR will appear ahead of the CMEs in HI-2 field-of-view \citep[]{Odstrcil:2009}. The CMEs are found to merge around 06:00UT on January 26. One of the main differences with the simulation of \citet{Lugaz:2009b} is the direction of propagation of the CMEs. In their simulations, \citet{Odstrcil:2009} initiate the CMEs on the eastern limb as seen from Earth. In \citet{Lugaz:2009b}, the CMEs are initiated in active region 10940 which was about 20$^\circ$ behind the eastern limb at the time of the first eruption and 10$^\circ$ at the time of the second eruption. The start time of the ejection was also different between these two simulations. In \citet{Odstrcil:2009}, the CMEs are initiated at 22~$R_\odot$ at the time corresponding to the onset time of the CMEs in LASCO. In \citet{Lugaz:2009b}, the time of the initiation of the CMEs was the time of first observation in COR1. 

The same events were reported in \citet{Webb:2009} with a simulation with the Hakamada-Akasofu-Fry Version 2 kinematic model \citep[HAFV2, see][]{Hakamada:1982, Fry:2001}. The steady-state solar wind is reconstructed using the Wang-Sheeley-Arge model with velocity pulses to initiate the transients. The initial velocities are determined from the fit to the velocities derived from LASCO data and the duration of the pulse from the duration of the GOES X-ray profile. This model predicts that the two shocks have merged by 12:00UT on January 26. 

These three heliospheric simulations --\citet{Lugaz:2009b}, \citet{Odstrcil:2009} and \citet{Webb:2009}-- are based on different numerical models and they use 3 different ways to initiate the CMEs: at the solar surface with an out-of-equilibrium flux-rope \citep[with the SMWF in][]{Lugaz:2009b}, at 2.5~$R_\odot$ with a velocity pulse \citep[with HAFV2 in][]{Webb:2009}, at 0.1~AU with a non-magnetic cone model \citep[with ENLIL in][]{Odstrcil:2009}. They all predict that the 2 CMEs are still distinct at 00:00UT on January 26 and have merged by 12:00UT. Both \citet{Lugaz:2008b} and \citet{Odstrcil:2009} show that one of the fronts observed in HI-2 is caused by a CIR or a dense stream, predicted to be behind the CMEs \citep[]{Lugaz:2009b} or ahead of them \citep[]{Odstrcil:2009}. However, in \citet{Lugaz:2009b}, we also tried to explain the origin of another front observed in HI-2 at times when all three numerical models predict the 2 CMEs have already merged (after 12:00UT on January 26). We believe that the first bright front observed in HI-2 corresponds to the flanks of the first CME. The flanks of the two CMEs are distinct on January 26, although their noses have merged (see bottom left and bottom middle panels of Figure 5). Because the CMEs originated from the limb, their flanks directly intersect the Thomson sphere and have a large contribution to the Thomson scattered signal. There are two main reasons why the noses and the flanks of the CMEs do not merge at the same time. First, the 2 CMEs propagate about 10$^\circ$ away from each other due to solar rotation. Second, the background solar wind is perturbed by the passage of the first CME, creating a more uniform background into which the second CME propagates. Consequently, the second CME flattens less than the first one (see bottom middle panel of Figure~5) and the distance between the CMEs' flanks is larger than that between the noses. 

Overall, the study of the January 24 and 25, 2007 CMEs by three different groups show the importance of numerical studies in support of SECCHI observations, especially for complex events, to correctly identify the bright fronts observed and also, to learn more about CME-CME interaction. It is indeed impossible to understand the complex interaction of ICMEs without using a physics-based numerical code, especially in this case where there was a data gap in the observations.

\section{Testing Analyses Methods}

Magnetic clouds observed at Earth have been associated with CMEs observed in the corona since the 1980s \citep[]{Burlaga:1982}. Associating complex structures observed in-situ with CMEs is not always as straightforward. It is also not always clear from observations only how the 3-part structure of some CMEs (front, cavity, core) relates to the typical shock-sheath-magnetic cloud from in-situ measurements \citep[see for example discussion in][]{Riley:2008}. 
Numerical simulations of ICMEs are particularly important to study the origin of the structures observed by in situ instruments on-board satellites. In \citet{Manchester:2004b}, the authors compare the results at 1~AU of a 3-D MHD simulation initiated with a flux rope \citep[]{Gibson:2000} with a typical three-part in-situ event (shock, sheath, magnetic cloud). They find good agreement between the simulation and the observation, which validates the results of the 3-D simulation. Moreover, synthetic line-of-sight images of the flux rope in the corona reveal that the flux rope is associated with the core and cavity of the CME, a result confirmed in a more recent study using a different initiation mechanism and solar wind model \citep[]{Riley:2008}. This type of work has enabled scientists to associate the bright front of CMEs with piled-up mass and compressed material, corresponding approximatively to the dense sheath observed in-situ prior to some ICMEs. 

In \citet{Riley:2004}, the authors use a numerical simulation to blindly  test different techniques to reconstruct magnetic clouds from in-situ measurements. They initiate  a CME at the solar surface via the ``flux cancellation'' method \citep[]{Riley:2003} and follow the erupting flux rope all the way to 1~AU, where they extract a synthetic satellite measurement. Different magnetic cloud reconstruction methods are then tested. Numerical simulations can be used as a sort of controlled laboratory environment, in the following manner. For the study of \citet{Riley:2004}, the magnetic topology of the magnetic cloud is {\em exactly} known, and the different reconstruction methods (force-free, Grad-Shafranov) can be rigorously tested, in particular with respect to the underlying assumptions.
\begin{figure}[ht*]
\begin{center}
\includegraphics*[width=6.5cm]{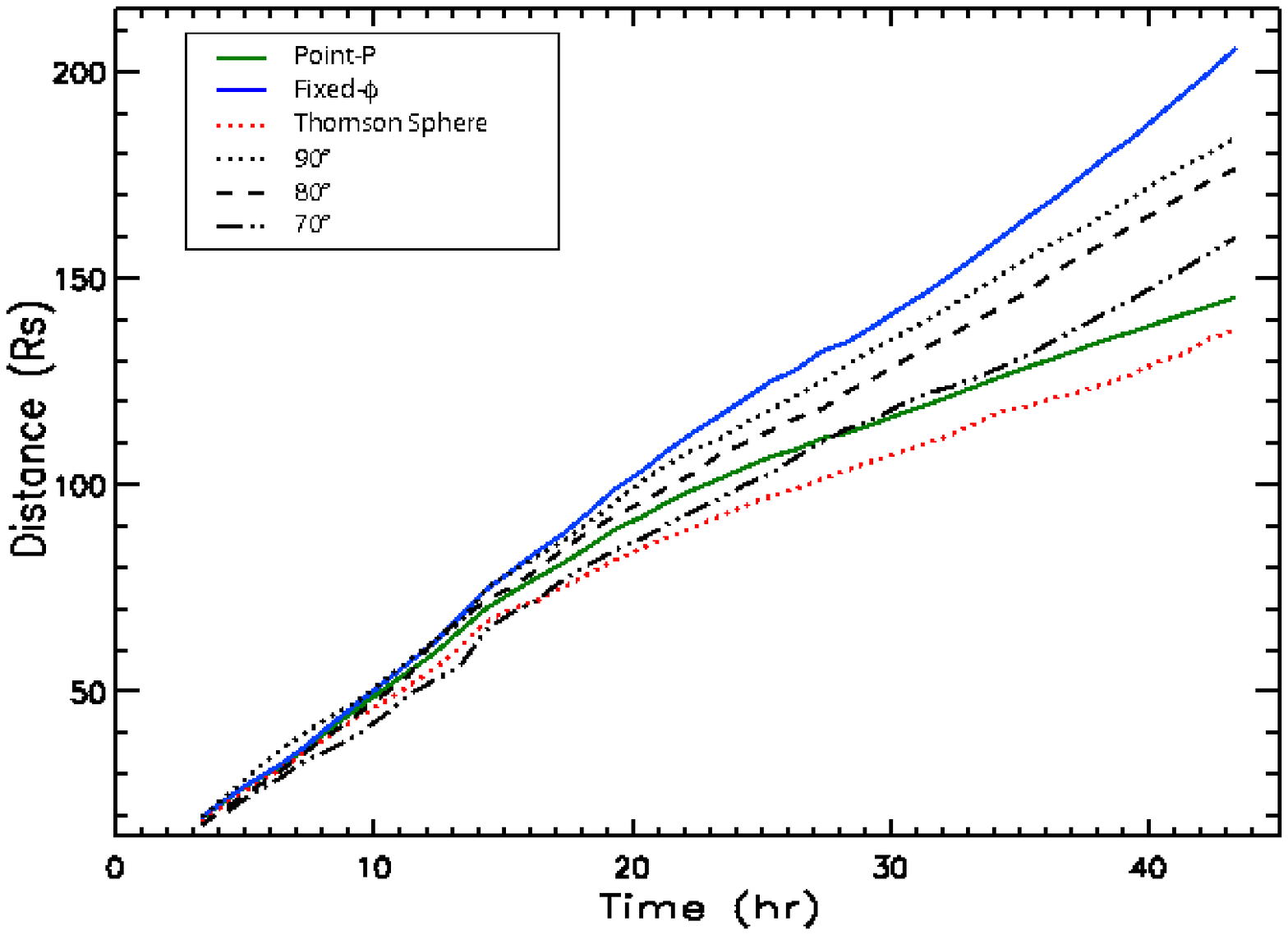}
\includegraphics*[width=6.5cm]{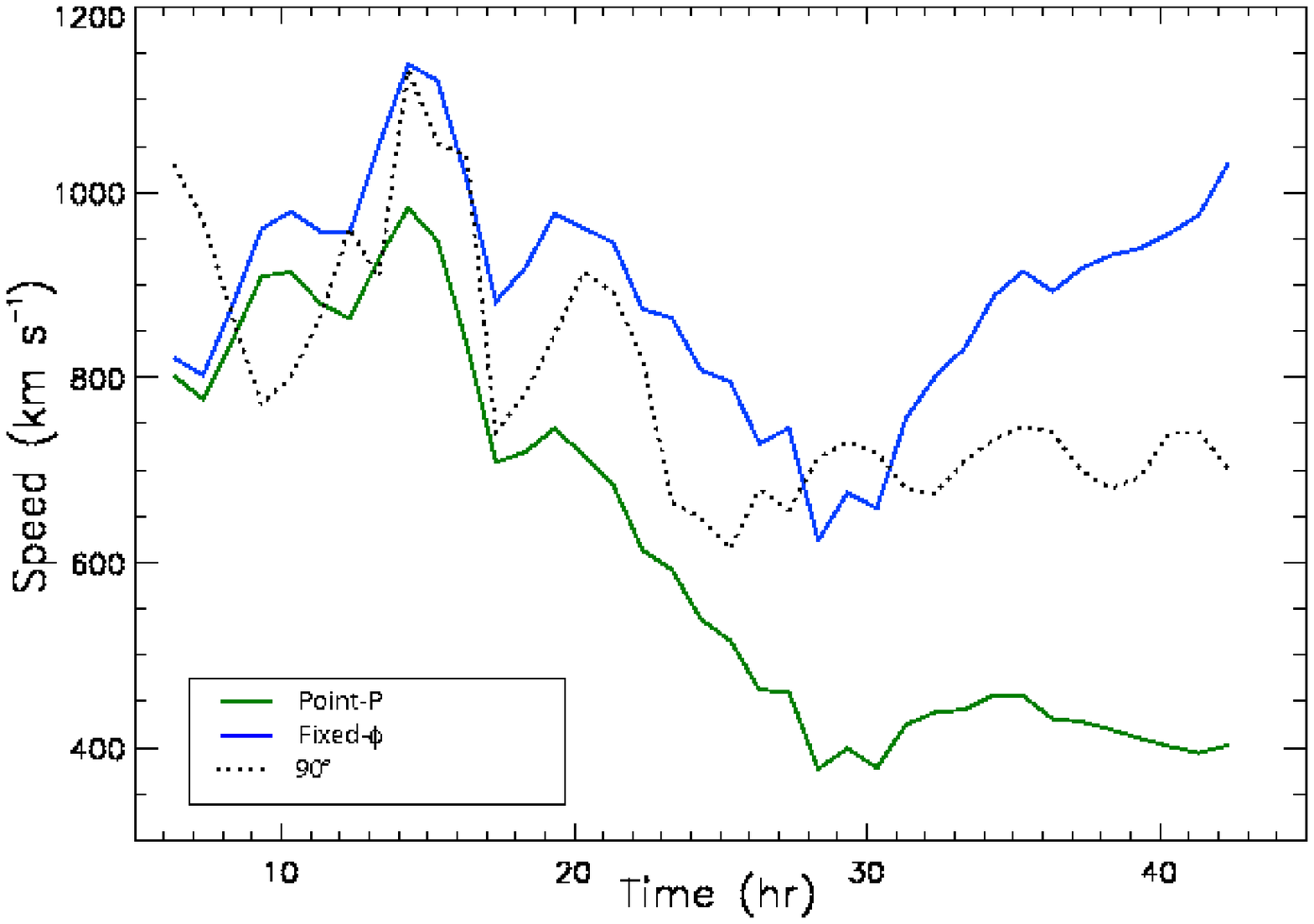}
\end{center}
\caption{Position ({\it top}) and speed ({\it bottom}) of the January 25, 2007 ICME at position angle 90$^\circ$ from the simulation along different radial trajectories, and, as derived from the synthetic SECCHI images with two different methods.}
\label{fig:HM}
\end{figure}
Theoretical error analyses can be performed to quantify the accuracy of the reconstruction: for example, by choosing slightly different boundaries for the ejecta or a slightly different axes. However, it is nearly impossible to assess from observations only the errors associated with the assumptions of the methods, for example the error associated with the assumption of force-free or hydrostatic equilibrium. \citet{Riley:2004} found that, when the spacecraft intersects the magnetic cloud off-axis, the resulting reconstruction may not identify this ejecta as a magnetic cloud. This is an important result because only about 30-40$\%$ of ejecta observed at 1~AU are catalogued as magnetic clouds \citep[]{Burlaga:2003}. From this study, it is now recognized that some of the remaining 60-70$\%$ are also magnetic clouds, but they are not recognized as such because of the large impact parameter \citep[]{Riley:2006}. This is the case, for example, when the flank of a magnetic cloud passes over the spacecraft. In this case, some of the typical properties of magnetic clouds are not observed in the in-situ measurements. Other non-magnetic cloud ejecta are, however, intrinsically so and we discuss them in section~5. 

In \citet{Lugaz:2009c}, we take a similar approach to test the two main techniques used to derive CME positions from elongation angle measurements made by SECCHI or SMEI. This work is a continuation of the study by \citet{Webb:2009} and based on the numerical investigation in \citet{Lugaz:2009b}. We analyze the simulated data from synthetic SECCHI images with two approximations: the Fixed-$\phi$ approximation \citep[]{Kahler:2005}, where the CME is assumed to be a single point propagating radially outward and the Point-P approximation \citep[]{Vourlidas:2006}, where the CME is assumed to be a spherical shell centered at the Sun.  Because the analysis is based on a simulated CME, the position of the CME in the three-dimensional domain is known exactly. We compare the position derived from the synthetic line-of-sight images with the position of the nose of the CME in the simulation. We also calculate the velocity from the reconstructed position and compare it with the CME speed from the simulation. This test of the approximations is particularly interesting to do for the January 25, 2007 CME because its speed varied significantly due to the interaction with the January 24 CME. We find (see Figure~\ref{fig:HM}) that the two approximations work well up to about 0.5~AU (here, corresponding to about 20-25 hours) with errors in position of less than 10$\%$. Beyond this distance, the Fixed-$\phi$ approximation over-estimates the CME position and predicts a CME acceleration (see right panel of Figure~\ref{fig:HM}). On contrary, the Point-P approximation under-estimates the CME position and predicts that the CME has decelerated down to the solar wind speed. These results do not reflect the 3-D simulation (see right panel of Figure~\ref{fig:HM}) and the two approximations should not be relied upon at large heliospheric distances. Knowing precisely the position and kinematics of CMEs is important to test different drag models as done in \citet{Webb:2009} and to improve space weather forecasting. 

\section{Magnetic Clouds and Complex Ejecta}

A number of recent studies have focused on non-cloud ejecta, which are thought to be of three origins: (1) magnetic clouds (MCs) which are not recognized as such because they are observed off-axis (see previous section), (2) ejecta, which are originally MCs but that lose the characteristics of MCs due to interaction during their propagation, and, (3) ejecta, which are not initially MC-like. In this section, we review recent works on the last two types of ejecta. In-situ measurement at 1 AU provide all what is known about these ejecta. Therefore, numerical simulations are an invaluable tool to study the 3-D magnetic topology of ejecta and the possible loss of the MC characteristics during propagation. 

\begin{figure*}[ht*]
\begin{center}
\includegraphics*[width=13cm]{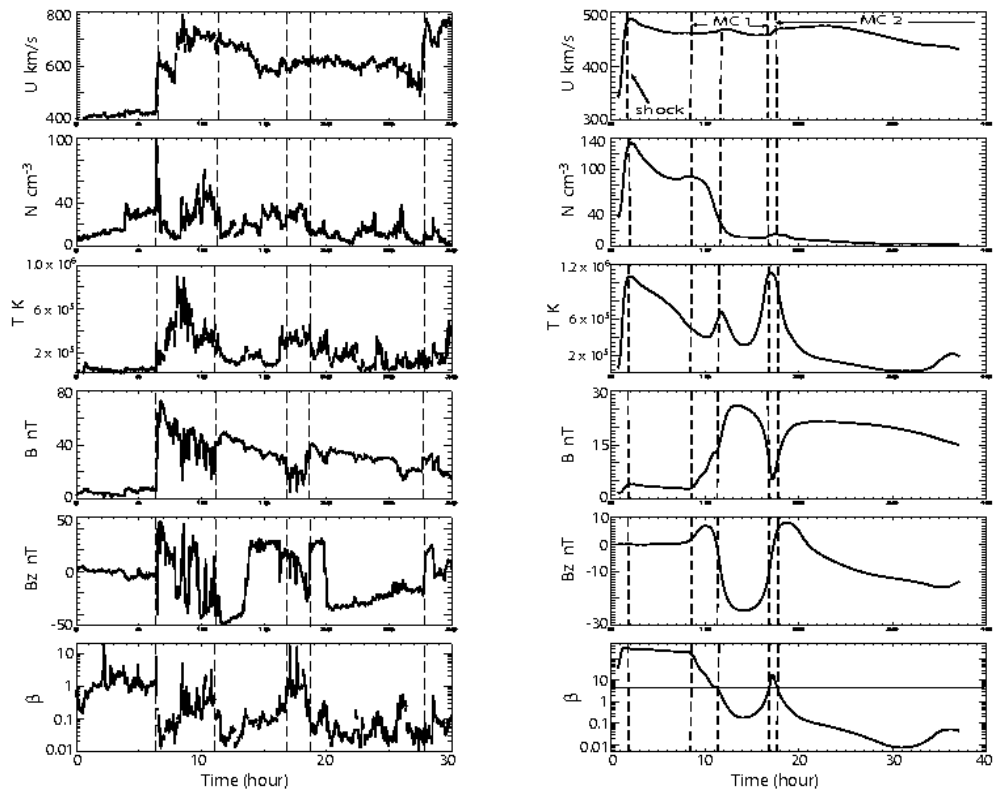}
\end{center}
\caption{{\it Left}: {\it ACE} measurements of multiple-magnetic cloud event in March-April 2001 as reported for example in \citet{Wang:2003b} The vertical lines show the position of the shock, the beginning and the end of the first cloud and the beginning and the end of the second cloud. {\it Right}: Results of our simulation of two successive CMEs as observed by a satellite at 1~AU as reported in \citet{Lugaz:2005b}. The vertical lines show the position of the shock wave preceding the magnetic clouds, the beginning of the first cloud, a possible contact discontinuity inside the first cloud, the end of the first cloud and the beginning of the second cloud, from left to right respectively. Note the zone of enhanced $\beta$ between the two clouds in the observations and the simulation results.}
\label{fig:sat}
\end{figure*}

\subsection{CME-CME interaction}
With the improvement of coronagraphic observations and the presence of spacecraft measuring the solar wind in the outer heliosphere (Voyager, Ulysses), it was recognized in the 1990s that successive ICMEs can merge with each other and also merge with CIRs to form a complex ejecta. In the outer heliosphere, these complex ejecta are sometimes referred to as a Merged Interaction Regions \citep[MIRs, see][]{Burlaga:1997}. The same phenomenon also happens in the inner heliosphere, before the CMEs reach Earth. One of the simplest cases is when two or more ejections interact but retain their individual ``identities'' and form what is known as a multiple-magnetic cloud event \citep[]{Burlaga:2002, Wang:2002, Wang:2003b}. More complex cases happen when it becomes nearly impossible to determine the solar origin of the complex ejecta observed in-situ. In this case, the characteristics of magnetic clouds are lost \citep[]{Burlaga:2003, Wang:2004}. Numerical simulations are useful in both cases to determine the origin of the ejecta observed at Earth and the interaction processes which yield the observed complexity. It is possible to derive from coronagraphic observations the initial speed and delays between eruptions. From there, it is possible to infer (or guess) which eruptions might have had an impact on the formation of a complex ejecta \citep[e.g., see][]{Burlaga:2003}. However, it is more rigorous to follow, via  numerical simulations, successive CMEs from the solar surface to Earth's orbit, and to treat self-consistently their interaction. In the past five years, such simulations have been performed in 1.5-D by \citet{WuCC:2005} in 2.5-D by \citet{Odstrcil:2005} and \citet{Wu:2002} and in 3-D by \citet{Schmidt:2004, Lugaz:2005b, Hayashi:2006, Xiong:2006, Xiong:2007, Lugaz:2007, WuCC:2007b, Odstrcil:2009}. 

\begin{figure*}[t*]
\begin{center}
\includegraphics*[width=13.cm]{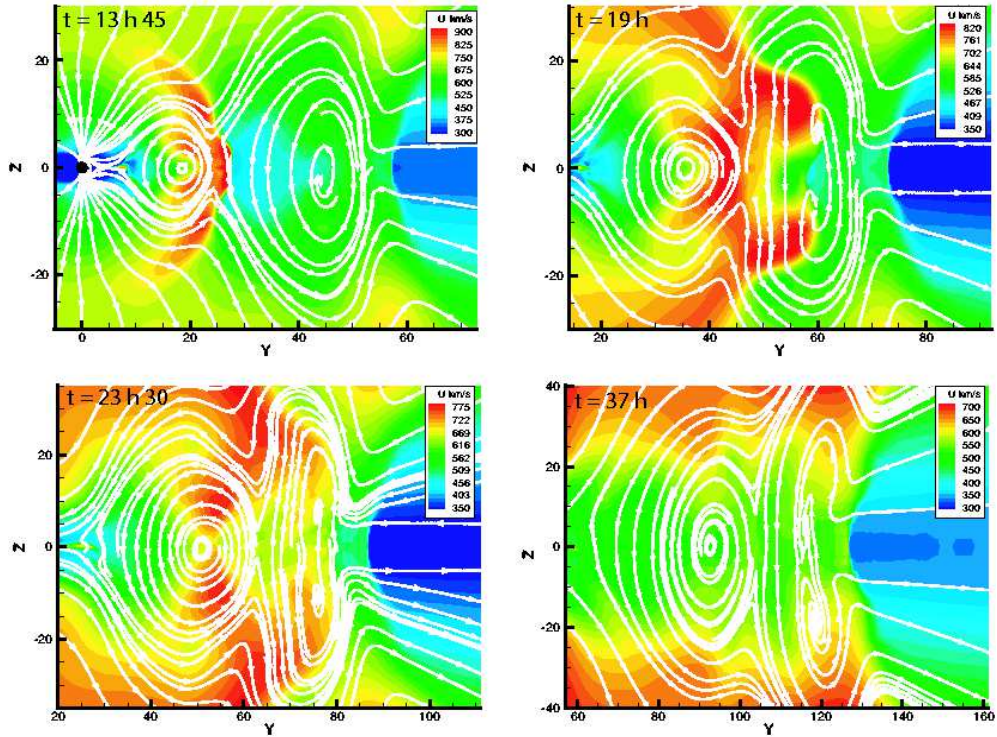}
\end{center}
\caption{Plasma speed in the $y-z$ plane at four different times during the interaction of the two CMEs. 
``Streamlines'' drawn in white illustrate the direction of the magnetic field
in the plane. 
{\it Top left}: t = 13 h 45, as the trailing shock enters the first cloud. 
{\it Top right}: t = 19 h, as the trailing shock has just passed the center of the first cloud. 
{\it Bottom left}: t = 23 h 30, as the trailing shock enters the dense sheath of plasma associated with the leading shock.
{\it Bottom right} t = 37 h. Note the uniformization of the speed in the two clouds at this time.}
\label{fig:speed}
\end{figure*}

Multiple-magnetic cloud events at 1~AU have the following characteristics: (1) a uniform speed profile, (2) a shorter duration of the first cloud compare to the second one, and, (3) a region of high $\beta$ between the clouds \citep[]{Wang:2003b}. The uniform speed appears to be associated with the ICME-ICME interaction, because the second ICME, in general, caught up with the first one and is expected to have a faster initial speed. It has been proposed that this uniform speed is due to the momentum exchange associated with an elastic collision between the ICMEs, or with an inelastic collision \citep[as proposed by][]{Farrugia:2004} or with the acceleration associated with the passage of a shock wave inside the first MC. These three mechanisms can also explain the shorter duration of the first cloud. Without numerical simulation, it is impossible to rule out any of these three mechanisms. 
In \citet{Lugaz:2005b}, we followed the interaction of two successive CMEs and we were able to reproduce and explain most of the features of multiple-magnetic cloud events (see Figure~\ref{fig:sat}): the uniform speed profile, the high values of the proton density and temperature in the sheath preceding the magnetic clouds and the existence of a region of high $\beta$ between the two clouds. We found that the speed of the event was relatively uniform, although the second CME caught up with the first one. This is due to a combination of two factors. First, the shock driven by the faster, overtaking CME propagates through the first CME and accelerates it, it results in the speed of the two CMEs to be comparable. After the shock wave has already compressed the back of the first ejecta,  the two ejecta collide which prevents the compressed first CME to expand and  to slow down (see Figure~\ref{fig:speed}). We also confirmed that the region of high-$\beta$ between the two eruptions is associated with magnetic reconnection between the two clouds. The magnetic energy is transformed into thermal energy, resulting in the enhanced $\beta$. Last, we found that shock-shock merging may result in a very dense sheath preceding the multiple-magnetic cloud event. It was proposed by \citet{Farrugia:2006} that these extremely high densities can fill up the plasmaphere and it may be an essential pre-conditionment before the arrival of the large southward interplanetary magnetic field found in the ejecta to explain the development of an intense geomagnetic storm. In \citet{Wang:2003a}, \citet{Xiong:2006} and in \citet{Lugaz:2008a}, it was found that the passage of the overtaking shock wave inside the overtaken magnetic cloud indeed results in very large values of the magnetic field. These simulations help explain why instances of CME-CME interaction are associated with some of the longest and largest geomagnetic storms \citep[]{Xie:2006}.

\subsection{Non-Twisted Ejecta}

\begin{figure*}[t*]
\begin{center}
\includegraphics*[width=6.5cm]{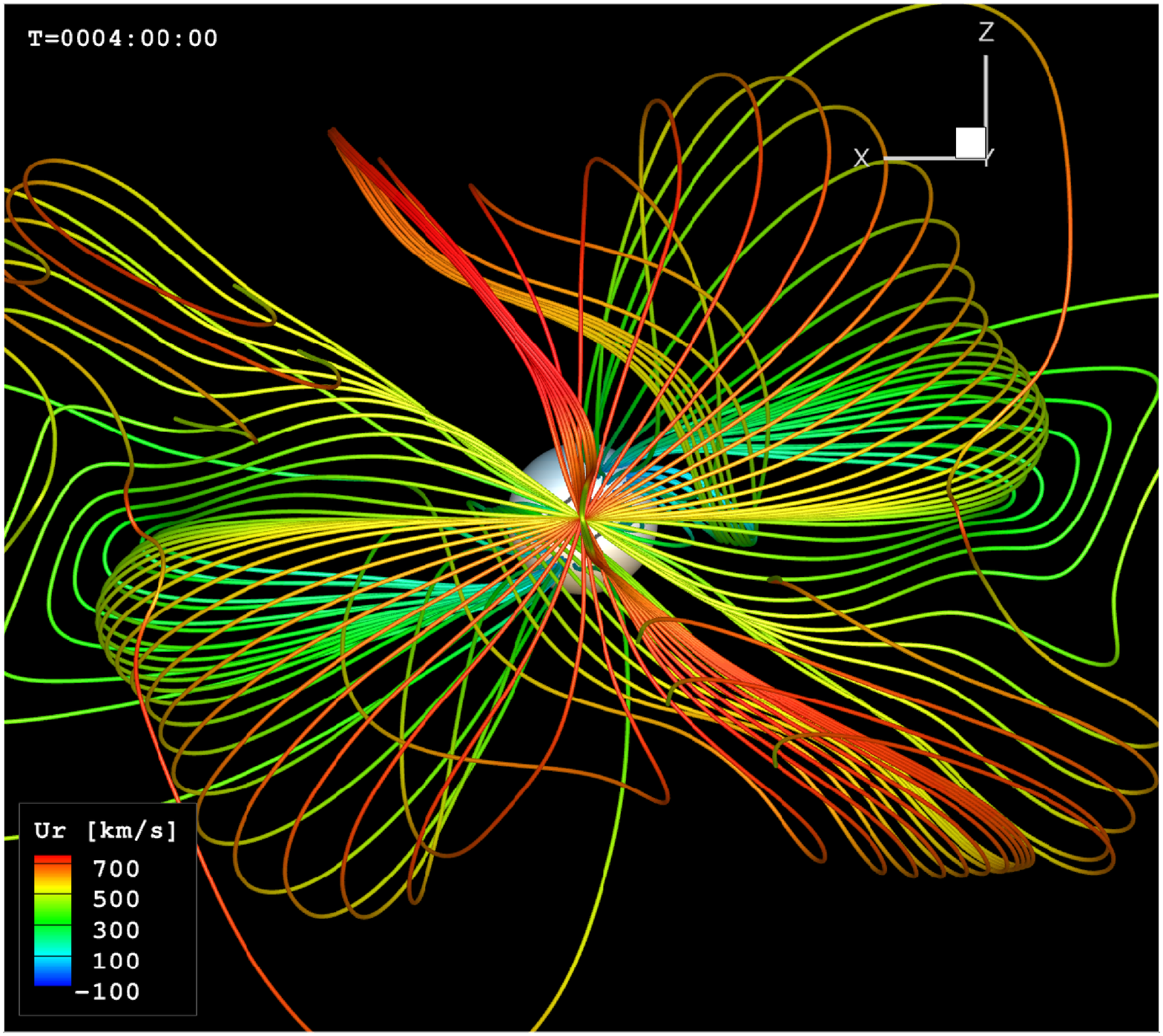}
\includegraphics*[width=7cm]{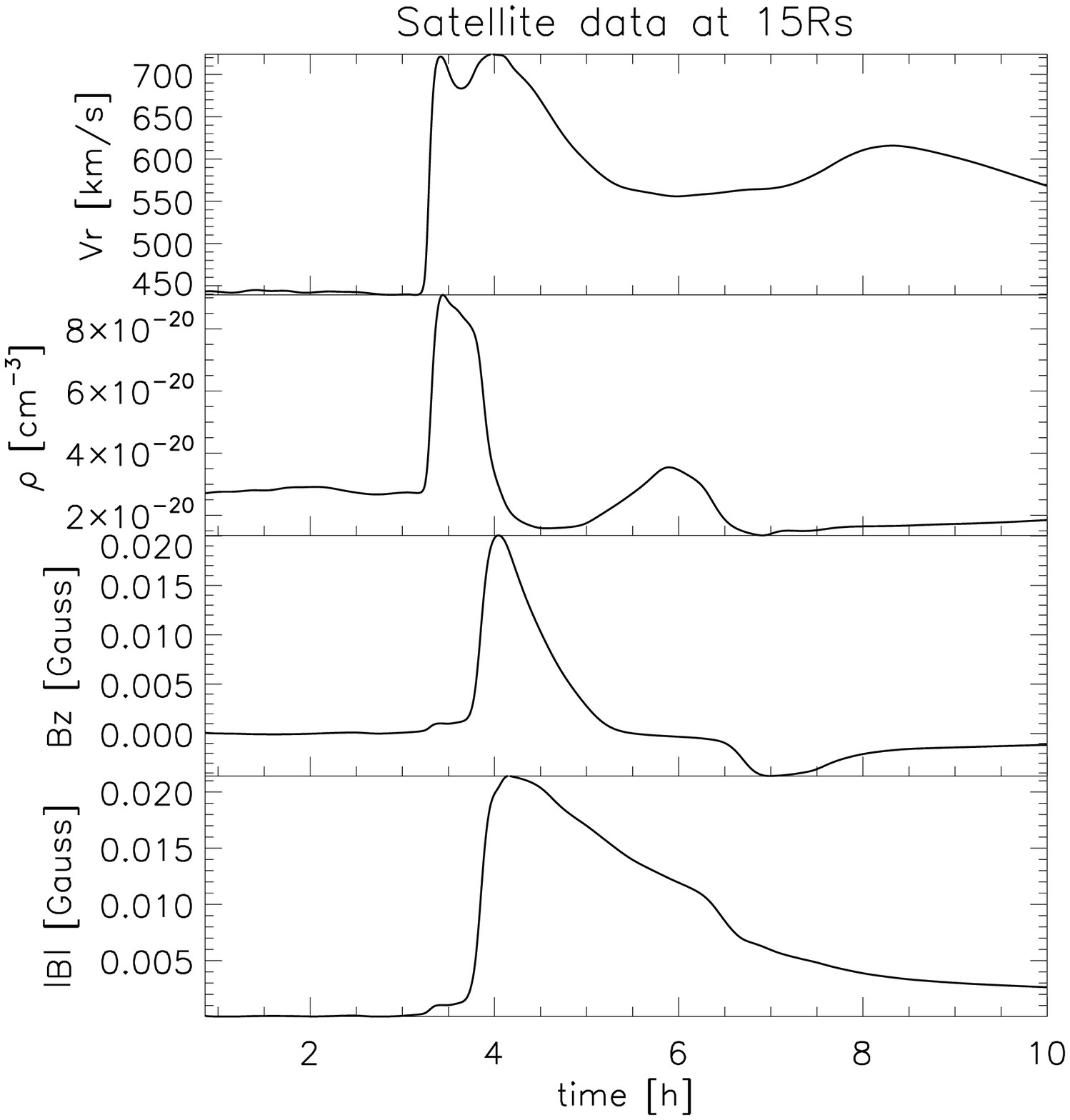}
\end{center}
\caption{{\it Left}: Simulation results at t = 4 hrs showing a non-twisted ejecta resulting from shearing motions (model of \citet{Roussev:2007}) applied to a dipolar active region. {\it Right}: Simulated satellite observations for this ejecta taken by an hypothetical spacecraft at 15~$R_\odot$ showing characteristics associated with a magnetic cloud and usually explained by a twisted flux rope ({\it from \citet{Jacobs:2009}}).}
\label{fig:last}
\end{figure*}

Other recent simulation efforts have focused on the existence of non-twisted ejecta. Magnetic clouds are commonly identified with twisted flux ropes. Some in-situ measurements of magnetic clouds appear to be in contradiction with this view. For example, in \citet{Moestl:2009a}, the authors find that the amount of twist is maximal about 1/2 to 2/3rd of the way inside the ejecta and not on its outside, and the minimum is not on the central axis of the cloud but near the outer boundary. In \citet{Moestl:2009c}, the authors find a near constant twist in all the field lines of the magnetic cloud. The model of \citet{Roussev:2007} attempts to mimic some of the shearing motions observed on the solar surface and relies on a realistic and complex magnetic topology using magnetogram data as input. This model was applied to the case studies of the April 21, 2002 CMEs \citep[]{Roussev:2007} as well as two successive CMEs in August 22 and 24, 2002 \citep[]{Lugaz:2009a, Lugaz:2010}. It was found that the CMEs do not develop into a twisted flux rope but, that, due to a series of reconnection events, the ejected magnetic field lines exhibit writhe. Individual field lines are not twisted but have different orientations from one another, so that two consecutive field lines have a slight angle between themselves \citep[]{Torok:2010}. The global aspect of the ejection exhibits smooth rotation of magnetic field lines without much twist (see Figure~\ref{fig:last}).
The model was modified and applied to simpler magnetic field topologies in \citet{Jacobs:2009}. The authors studied the shearing of a dipole embedded into a dipolar active region and into a quadrupolar active region. As the writhed magnetic field lines propagate past an imaginary spacecraft, the satellite would observe a rotation of the magnetic field vector, reminiscent of a magnetic cloud, as can be seen in the bottom panel of Figure~\ref{fig:last}. Future work will test whether reconstruction codes such as the one by \citet{Moestl:2009a} would mistake synthetic data produced from simulations without twist for a twisted flux rope. If this is the case, the general paradigm of magnetic cloud as a twisted flux rope would have to be revised.

\section{Discussions and Conclusion}

In this article, we have reviewed some of the recent simulations of ICME propagation and ICME-ICME interaction in the interplanetary space. We have discussed how simulations can be used in support of heliospheric missions (for example {\it STEREO}), and to study physical questions such as shock-ICME interaction and heliospheric drag. The two main types of heliopsheric MHD models have boundaries at the solar surface or beyond the critical point (typically 0.1~AU), and they can be used to study with different level of details ICME propagation and interaction. When these two different types of models, as well as a kinematic model, are used to simulate the same observed event (here, the January 24-25, 2007 CMEs), they predict a similar behavior overall, but the differences in the details have important consequences. For example, the propagation of a shock wave inside a magnetic cloud has been found to greatly change the properties of the magnetic cloud as well as the properties of the shock wave \citep[]{Lugaz:2005b}. Kinematics simulations and MHD simulations which do not use a magnetized ejecta are not expected to correctly reproduce the change in speed of a shock wave inside a magnetic cloud. This might be the reason why the models of \citet{Fry:2001} and \citet{Odstrcil:2009} do not predict the CME-CME merging at the right heliocentric distance for this event.

Numerical models have also been used to make predictions of physical phenomena in the heliosphere. It is for example the case for the change of indentation of ICMEs during solar minimum conditions as they propagate close to the current sheet \citep[]{Riley:2001, Manchester:2004b}. While there have been some studies, which appeared to confirm these findings \citep[]{Kahler:2007, Liu:2008, Savani:2010}, almost all CMEs observed by SECCHI do not show any sign of an indentation change. We suspect that the greater number of CIRs during the current solar minimum \citep[]{Mason:2009} compared to the previous one, is one potential cause for the discrepancy between simulations and observations. Simulations have also been used to study the nature of non-magnetic cloud ejecta, which are from 3 origins: (1) magnetic cloud-like ejecta which are not recognized as such because they are seen off-axis as shown in \citet{Riley:2004}, (2) complex ejecta resulting from the interaction of successive CMEs \citep[]{Lugaz:2007}, and, (3) ejecta which are intrinsically non-twisted flux ropes, but which may have some of the characteristics of magnetic clouds due to writhe \citep[]{Jacobs:2009}. 

Most heliospheric simulations of ICMEs use relatively simple models, from pressure or velocity pulses to out-of-equilibrium flux ropes initiated at the solar surface. By comparison, simulations of CMEs in the corona use more complex and realistic models involving shearing motions and flux emergence among other processes \citep[]{Mikic:1995, Antiochos:1999, Amari:2003a, Amari:2004, Manchester:2007, Roussev:2007, Lynch:2008, Holst:2009, Aulanier:2010}. Most of these simulations follow the propagation of the CMEs up to few tens of solar radii at most and do not always have a realistic solar wind and/or coronal magnetic field. With the continual improvement of computational resources and the existence of interplanetary measurements to compare simulations to, it is now important to do heliospheric simulations of ICMEs using realistic models of CME initiation, to test these different models and study their consequences. Only in the past few years, they have been some attempts to do this, for example in \citet{Lugaz:2010} with the August 24, 2002 CME and the model of \citet{Roussev:2007}. On the other hand, simpler models are and will still be useful for space weather forecasting \citep[]{Fry:2003, Takta:2009}. In particular, they could be used to provide ensemble forecasting for space weather in a way comparable to what is done now with weather and climate modeling. Finally, combining heliospheric observations and the modeling of real events is expected to improve our knowledge of CME-CME interaction. Recent studies have explained why CME-CME interactions are associated with some of the largest geomagnetic storms: the compression of the magnetic field inside magnetic clouds and the formation of dense sheaths. It is important to study individual events to understand when and how these two phenomena happen. We believe further modeling of real events will reveal the importance of studying individual events on case-by-case basis not only at the Sun, as stated in \citet{Roussev:2007} but also in the heliosphere. We expect more progresses to be achieved in these three directions (testing of coronal models, ensemble forecasting of space weather and better understanding of individual events, in particular complex ones) during solar cycle 24. 

\section{Acknowledgments}
The authors would like to thank the anonymous reviewers  for their useful comments as well as Nada Alhaddad, Robin Colaninno, Chris Davis, Cooper Downs, Carla Jacobs, Ward Manchester, Dusan Odstrcil,  Angelos Vourlidas and Dave Webb for useful discussions. The research for this paper was supported by the following grants: NSF grants ATM-0639335 and ATM-0819653 as well as NASA grants NNX-07AC13G and NNX-08AQ16G.  Simulation results were obtained using the Space Weather Modeling Framework, developed by the Center for Space Environment Modeling, at the University of Michigan with funding support from NASA ESS, NASA ESTO-CT, NSF KDI, and DoD MURI. The SECCHI data are produced by an international consortium of a
  Naval Research Laboratory, Lockheed
  Martin Solar and Astrophysics Lab, and NASA Goddard Space Flight
  Center (USA), Rutherford Appleton Laboratory, and University of
  Birmingham (UK), Max-Planck-Institut f{\"u}r Sonnensystemforschung
  (Germany), Centre Spatiale de Liege (Belgium), Institut d'Optique
  Th{\'e}orique et Appliqu{\'e}e, and Institut d'Astrophysique
  Spatiale (France).






\end{document}